%% file: main.tex
\documentclass[a4paper,10pt]{article}
\pdfoutput=1 % pdflatex

\usepackage{jheppub}

\usepackage[T1]{fontenc}
\usepackage[normalem]{ulem} % add this in the preamble
\usepackage{float}
\usepackage{comment}
\usepackage{subcaption}
\allowdisplaybreaks
\usepackage{amsmath,amssymb,xcolor,graphicx,subcaption}
\usepackage{tikz}
\usetikzlibrary{decorations.pathmorphing, arrows.meta}
\definecolor{softgreen}{RGB}{164,210,170}
\tikzset{
  quark/.style   = {line width=1.2pt},
  flowarrow/.style = {-{Stealth[length=4pt,width=5pt]}, line width=1.0pt},
  gluon/.style   = {decorate, decoration={coil, segment length=3.2pt, amplitude=2.6pt}, line width=1.1pt},
}

\newcommand{\kus}{k_{\rm us}}
\newcommand{\kapus}{\kappa_{\rm us}}
\newcommand{\etaus}{\eta_{\rm us}}
\newcommand{\phius}{\phi_{\rm us}}

\author[a]{Andrea Banfi,}
\author[b]{Basem Kamal El-Menoufi,}

\affiliation[a]{Department of Physics and Astronomy, University of Sussex,\\Sussex House, Brighton, BN1 9RH, U.K.}
\affiliation[b]{School of Physics \& Astronomy, Monash University,\\Wellington Rd, Clayton VIC-3800, Australia}

\emailAdd{a.banfi@sussex.ac.uk}
\emailAdd{basem.el-menoufi@monash.edu}

\title{Characterising hadronisation across the phase space}

\abstract{Leading hadronisation corrections to two-jet global event shapes amount to a shift in the corresponding perturbative distributions. It has been recently established that this shift depends significantly on the value of the considered event shape. These analyses consider perturbative configurations with only three partons emitting an ultra-soft non-perturbative gluon. These are dominant in the three-jet region. However, multiple soft and/or collinear emissions need to be considered to accurately describe event shape distributions near their Sudakov peak. In this region, non-perturbative shifts are usually computed by considering only two hard emitters. In this paper, we find that this approximation misses an important correction to the shift due to the emission of an additional soft wide-angle gluon. We then compute its contribution, and embed it in a general treatment for the shift that is valid in both the two-jet and three-jet regions.}

\begin{document} 
	
	\maketitle
	\flushbottom
%%%%%%%%%%%%%%%%%%%%%%%%%%%%%%%%%%%%%%%%%%%%%%%%%%%%%%%
% Sections
%%%%%%%%%%%%%%%%%%%%%%%%%%%%%%%%%%%%%%%%%%%%%%%%%%%%%%%
\input{Introduction}

\input{basics}
\input{recovery}
\input{resum_dist}
\input{matching}
\input{conclusions}
\input{app_ABC}
%\input{Recovery of mean value, comment on anomalous dimension--}

%\clearpage
\bibliographystyle{JHEP}
\bibliography{hadronisation}

\section*{}
%\clearpage
%\addcontentsline{toc}{section}
%{Index}
%\printindex

\end{document}

%% file: introduction.tex
\section{Introduction}
	The analysis of increasingly precise measurements from high-energy colliders demands commensurate advances in the theoretical accuracy of Standard Model (SM) predictions. Despite substantial progress in perturbative Quantum Chromodynamics (QCD) calculations, significant foundational work remains on multiple fronts. Two challenges are particularly pressing. First is the need for a deeper understanding of the all-order structure of collider observables, which are almost invariably dominated by perturbative QCD dynamics, see e.g. \cite{Luisoni:2015xha}. Second, as perturbative (PT) precision reaches higher orders, the unavoidable fact that collider measurements involve non-perturbative (NP) hadrons rather than the quarks and gluons of perturbative QCD necessitates a detailed treatment of hadronisation effects in such observables, see e.g. \cite{Altarelli:1995kz, Beneke:1998ui}.

    A common strategy to quantify the impact of non-perturbative physics is the use of phenomenological models that describe the transition from partons to hadrons, i.e.\ hadronisation models used extensively in Monte Carlo event generators \cite{Andersson:1983ia, Webber:1983if}. Although successful in practice, these models rely on parameters which are tuned to data. As such, this framework provides limited analytical insight and face long-standing challenges in being consistently combined with state-of-the-art perturbative calculations. An alternative route relies on analytic insights drawn from the renormalon divergence of QCD perturbation theory at high orders, reflected in the factorial growth of perturbative coefficients \cite{Altarelli:1995kz, Beneke:1998ui}. 
    %Infrared renormalon divergences manifest as poles in the complex Borel plane, prompting an ambiguity in Borel summation which is remedied by augmenting the perturbative prediction by non-perturbative corrections. 
    %
    This perspective offers a potential window into the underlying structure of non-perturbative effects and thus into the analytical nature of hadronisation.
    Non-perturbative corrections generally take the form $\left(\Lambda_{\rm QCD}/Q\right)^p$, where $Q$ denotes the hard scale of the perturbative series, and the exponent $p$ depends on the observable under consideration. The numerically dominant class of non-perturbative corrections corresponds to the linear case, $p=1$. Such corrections affect nearly all infrared-safe observables constructed from hadron momenta. First investigated in the mid-1990s~\cite{Dokshitzer:1995zt,Akhoury:1995fb,Nason:1995np,Dokshitzer:1995qm,Dokshitzer:1997ew,Dokshitzer:1997iz,Dasgupta:1999mb,Korchemsky:1999kt}, they remain, from a phenomenological standpoint, central to ongoing determinations of the strong coupling constant from event-shape data~\cite{Banfi:2023mes,Bell:2023dqs,Nason:2025qbx}, where discrepancies of up to three standard deviations persist relative to the current world average~\cite{Gehrmann:2012sc,Hoang:2015hka,PhysRevD.110.030001,Benitez:2024nav,Benitez:2025vsp}.

    One of the central insights of the 1990s was that the renormalon framework effectively reduces to studying how the emission of a single gluon with low transverse momentum modifies an observable. This gluon, often referred to as a {\em gluer}~\cite{Dokshitzer:1995qm}, is assumed to carry transverse momentum well below that of perturbative emissions. Consequently, a key assumption in analytic models is that the gluer’s contribution to an observable is subleading compared to perturbative effects. Within this framework, non-perturbative corrections to event-shape variables manifest as shifts in their distributions.

    In $e^+e^-$ annihilation, the earliest analyses examined the mean value of event shapes when a gluer is emitted from the final-state $q\bar{q}$ colour dipole \cite{Dokshitzer:1995zt}. In this simple setting, the modification of the mean value is straightforward to evaluate, since the event-shape variable vanishes at the Born level. Subsequent studies extended the gluer model to event-shape distributions, exploiting the rich experimental data to fit the strong coupling and the non-perturbative parameter simultaneously. Nevertheless, to properly characterise event-shape data, resummation of all-order logarithms becomes essential in appropriate regions of phase space. The gluer model has often been carried over under the assumption that it remains valid even in the presence of the soft-collinear ensemble of perturbative emissions responsible for next-to-leading logarithms (NLL).

    One of the pressing challenges of the gluer picture concerned the phenomenological soundness of the model. In practice, the non-perturbative shift was applied uniformly across phase space, extending well beyond the resummation region. At fixed-order, recent works showed that the gluer model can nevertheless be employed throughout phase space, provided that the gluer is consistently inserted into every colour dipole of the event \cite{Caola:2022vea}. This procedure requires a recoil scheme, which enforces overall momentum conservation and enables one to determine the modification of the observable when the gluer is included in the event. The main result of these studies is that the shift becomes strongly dependent on the observable's value. Additionally, the value of the shift in the three-jet region differs substantially from its value in the two-jet region. This happens in particular in the region around the peak of event-shape distributions, where events are two-jet-like, thus questioning the usefulness of calculations performed in the strict two-jet limit to describe experimental data \cite{Nason:2023asn}. 

    The question we wish to address in this work is whether it is possible to unify the treatment of NP corrections to event-shape distributions in the two- and three-jet regions. Given the technical nature of the paper, we introduce the problem and our main results here, leaving the details to individual sections. 

    To set the stage, we consider an event-shape variable $V$, a function of all final-state momenta, and its differential distribution as a function of the event shape's value $v$. Let us first analyse the origin of the departure of the value of the shift from the {\em strict} two-jet limit. In the two-jet limit, the starting point is a $q\bar q$ dipole accompanied by arbitrarily many soft and/or collinear gluons. In particular, in this limit all existing calculations within the dispersive model consider only perturbative soft and collinear gluons.
    These emissions give a contribution of order $(\alpha_s L)^n$, with $L=\ln(1/v)$. To be more concrete, let us consider the thrust where from now onwards we use the notation $\zeta_V(v)$, introduced in ref.~\cite{Caola:2022vea}, to denote the NP shift for the event shape $V$, which is generally a function of $v$. To be more specific, we consider the shift for the thrust $T$ as a function of $\tau$, the value of $1\!-\!T$. At the lowest order in $\alpha_s$, the full NP shift for the thrust distribution is obtained from the quantity,
    \begin{equation}
    \label{eq:starting-point}
         \frac{\tau}{\sigma_0} \frac{d\sigma}{d\tau} \zeta_T(\tau) = 2\times 2\frac{\alpha_s}{\pi} C_F \ln\frac{1}{\tau} +\dots\qquad
        \frac{\tau}{\sigma_0}\frac{d\sigma}{d\tau} = \frac{\alpha_s}{\pi} C_F\left(2\ln\frac{1}{\tau}-\frac 32\right)+\dots
    \end{equation}
    where the dots represent subleading contributions. In the two-jet limit, $\tau\to 0$, we have $\zeta_T(\tau) \to 2$.  
    Using the expressions in eq.~\eqref{eq:starting-point}, we find:
    \begin{equation}
        \frac{\zeta_T(\tau)}{\zeta_T(0)} \simeq \frac{4\ln(1/\tau)+c}{4\ln(1/\tau)-3}\simeq 1+\frac{c+3}{4\ln(1/\tau)}\,,
    \end{equation}
    where the constant $c$ represents the subleading contribution in eq.~\eqref{eq:starting-point}. This ratio will always tend to one in the two-jet limit, but with large corrections, unless the constant $c$ is exactly equal to $-3$. However, as we have already remarked, all current formulations of NP corrections in the two-jet limit predict only the term of order $\alpha_s\ln(1/\tau)$ in eq.~\eqref{eq:starting-point}, but fails to capture contributions of order $\alpha_s$. In other words, there is no guarantee that $ \zeta_T(\tau) \,\tau d\sigma/\sigma_0 d\tau$ is proportional to the leading-order (LO) thrust distribution. Contributions of order $\alpha_s$ may originate from two mechanisms: the emission of a single hard–collinear gluon, or the emission of a soft wide–angle gluon that subsequently acts as a new emitter for the non–perturbative gluon.
    We find that:
    \begin{itemize}
        \item The hard–collinear contribution amounts to twice the hard–collinear constant ($-3$), a consequence of colour coherence: the ultra–soft gluon does not resolve the collinear pair from its parent emitter. As a result, this term in the shift is proportional to $C_F$.
        \item The emission of a soft wide–angle gluon furnishes two extra dipoles which act as sources for the non–perturbative gluon, thereby producing a novel $\mathcal{O}(\alpha_s)$ contribution proportional to $C_A$. This constant has been the subject of recent works in the context of NP corrections to mean values \cite{Dasgupta:2024znl,Farren-Colloty:2025amh}. 
    \end{itemize}

    We introduce a general framework to determine the soft–collinear constant and to compute the associated coefficient functions governing the departure from the $\tau \to 0$ limit. Up to first order in $\alpha_s$, the result reads: 
    \begin{align}
    \label{eq:end-point}
    \nonumber
        \frac{\tau}{\sigma} \frac{d\sigma}{d\tau} \zeta_T(\tau) = \frac{\alpha_s}{2\pi}\bigg[
        &C_F \,\zeta_T(0) \left(4\ln\frac{1}{\tau} - 3 \right) + C_F\, c_{q\bar q}(\tau)\\
        +&C_A\bigg( 16 (\ln 2-1) + \frac12 \left(c_{qg}(\tau)+c_{g\bar q}(\tau)-c_{q\bar q}(\tau)\right)\bigg)
        \bigg]\, ,
    \end{align}
    where $\sigma$ is the total cross section and the functions $c_{q\bar q}, c_{qg}$ and $c_{g\bar q}$ are suppressed by powers of $\tau$. 

    If we want to merge the expression of the NP shift across the two- and the three-jet regions, we need to supplement the result in eq.~\eqref{eq:end-point} with the effect of multiple soft and collinear gluons. At the level of the differential cross section, this gives:
    \begin{equation}
        \label{eq:dsigma/dtau-NLL}
        \frac{\tau}{\sigma}\frac{d\sigma}{d\tau} \simeq \frac{d\Sigma_{\rm NLL}}{d\tau}= R' \mathcal{F}(R')\, e^{-R(\tau)}\,,\qquad R'=-\tau \frac{dR}{d\tau}\simeq 2 \frac{\alpha_s}{\pi}
        C_F\ln\frac{1}{\tau}\, ,
    \end{equation}
    where $\mathcal F(R')$ a well-known finite function of $R'$. When only soft {\em and} collinear gluons emit the gluer, we have:
    \begin{equation}
    \label{eq:NPshift-collinear}
      \frac{\tau}{\sigma} \frac{d\sigma}{d\tau} \zeta_T(\tau) \simeq  \zeta_T(0) \times R' \mathcal{F}(R')\, e^{-R(\tau)} \, .
    \end{equation}
    We find that including PT configurations with a single {\em hard} and collinear gluon gives the same result as in eq.~\eqref{eq:NPshift-collinear}, provided the expression of $R'$ contains hard-collinear contributions as well. This is due to QCD coherence. However, the NP gluon can resolve an additional PT soft gluon at large angles. We find that we can distinguish two regimes, according to the size of the transverse momentum $k_t$ of the PT soft gluon:
    \begin{itemize}
    \item $ k_t\sim \tau Q$: this gives a finite contribution to the shift, proportional to $C_A \alpha_s(\tau Q)$, whose fixed-order expansion coincides with eq.~\eqref{eq:end-point}.
    \item $k_t \ll \tau Q$: this gives rise to a logarithmically enhanced contribution $\ln(\tau Q/\kappa_{\rm us})$, which is the first order expansion of an anomalous dimension for the NP shift, where $\kapus$ is the transverse momentum of the gluer and $\kapus \ll k_t$.
    \end{itemize}
    A detailed calculation gives an expression for the NP shift that is valid across the phase space:
    \begin{equation}\label{eq:shift_intro}
    \langle\delta (1 \!-\! T) \rangle = \frac{1}{Q} \left\langle \kapus  \exp\left[2 C_A \gamma \int_{\kapus}^Q \!\! \frac{dk_t}{k_t}\frac{\alpha_s(k_t)}{\pi}\right]\right\rangle_{\rm NP} \zeta_T(\tau) , \qquad \gamma \equiv 4 (\ln 2 - 1) \, ,
    \end{equation}
    where $\gamma$ is an anomalous dimension and $\langle \cdots \rangle_{\rm NP}$ is a non-perturbative average which will be defined in the sequel and defines the NP parameter. Additionally, we have:
    \begin{multline}\label{eq:hT-final}
    \frac{\tau}{\sigma} \frac{d\sigma}{d\tau}  \zeta_T(\tau) 
    \simeq \left(\frac{\alpha_s(Q)}{\alpha_s(\tau Q)}\right)^{\frac{\gamma }{\pi \beta_0 \zeta_T(0)} C_A}
    \bigg[  \tau \frac{d\Sigma_{\rm NLL}}{d\tau}\, \left(\zeta_T(0)
    + \frac{2 C_A \gamma\, \alpha_s(\tau Q)}{\pi}\, \left(\frac{1}{R'} + \frac{d}{dR'}\ln \mathcal{F}(R') + c_1^{\rm w.a.} \right)\right)\\
     + \frac{C_F \alpha_s(Q)}{2\pi}
   \left(
     c_{q\bar{q}}(\tau) + \frac{C_A}{2C_F} 
     \left(  c_{qg}(\tau) + c_{g\bar{q}}(\tau) - c_{q\bar{q}}(\tau)\right)
   \right) \bigg]\, ,
    \end{multline}
    with $c_1^{\rm w.a.}$ a calculable constant. The most important feature of our calculation is that we are still able to write the final expression of the shift, eq.~\eqref{eq:shift_intro}, as the product of a perturbatively calculable coefficient $\zeta_T(\tau)$ multiplying a genuinely non-perturbative quantity. This NP quantity, i.e. $\langle \cdots \rangle_{\rm NP}$, also appears in determining the NP shift to the thrust mean value. 

The paper is organised as follows. In section~\ref{sec:prelims}, we present the prerequisites to compute the NP shift for the thrust in the presence of three hard emitters. In section~\ref{sec:recovery} we recapitulate the results of ref.~\cite{Caola:2022vea}. %This gives its expression in the three-jet region, at order $\alpha_s$, which we recalculate in section~\ref{sec:qg2jet}. 
Then, in section~\ref{sect:2jextract}, we extract the two-jet limit of the shift starting from the $\mathcal{O}(\alpha_s)$ result. This gives rise to the result in eq.~\eqref{eq:end-point}, where the anomalous dimension $\gamma$ makes its first appearance. 
To smoothly interpolate between the 2- and 3-jet regions, in section~\ref{sec:shift-sc} we augment the fixed-order shift by including an arbitrary number of soft and collinear gluons. This leads to the interpretation of $\gamma$ as an anomalous dimension, and ultimately to the result in eq.~\eqref{eq:hT-final}. Finally, in section~\ref{sec:matching}, we match the full shift to fixed order and provide numerical comparison of $\zeta_T(\tau)$ to the strict two-jet and three-jet limits. Additional technical details are collected in the appendices.

%     \begin{figure}
%     \centering
%     \begin{tikzpicture}

%     %-------------------- (a) --------------------
%     \begin{scope}[xshift=0cm, yshift=0cm]
%     % label
%     \node[anchor=center] at (3.6,0.55) {(a)};

%     % parton line
%     \draw[quark] (3.2,0) -- (-3.2,0);

%     % little flow arrows (optional, direction cues)
%     \draw[flowarrow] (-1.6,0) -- (-2.5,0);
%     \draw[flowarrow] ( 1.6,0) -- ( 2.5,0);

%     % the "gluer"
%     \draw[gluon, red] (0,0) -- ++(0.55,0.75);
%     \end{scope}

%     \end{tikzpicture}
%     \end{figure}

%     \begin{tikzpicture}
%         %-------------------- (b) --------------------
% \begin{scope}[xshift=7.2cm, yshift=0cm]
%   % label
%   \node[anchor=west] at (-3.6,0.55) {(b)};

%   % soft band
%   \fill[softgreen!45] (-3.2,0.17) rectangle (3.2,-0.17);

%   % parton line
%   \draw[quark] (-3.2,0) -- (3.2,0);

%   % flow arrows
%   \draw[flowarrow] (-2.5,0) -- (-1.6,0);
%   \draw[flowarrow] ( 2.5,0) -- ( 1.6,0);

%   % emission point
%   \coordinate (x0) at (0.2,0);

%   % the soft "gluer" at the emission point
%   \draw[gluon, red] (x0) -- ++(0.45,0.6);

%   % pale wedges suggesting soft/collinear region
%   \foreach \ang/\len in {70/1.5, 95/1.35, 120/1.45}{
%     \path[softgreen!35] (x0)
%       -- ++({\ang-6}:0.18)
%       -- ++({\ang}:\len)
%       -- ++({\ang+6}:-0.18-\len)
%       -- cycle;
%   }

%   % a few harder gluon emissions
%   \foreach \ang/\len in {80/1.2, 105/1.35, 130/1.1}{
%     \draw[gluon] (x0) -- ++(\ang:\len);
%   }
% \end{scope}
%     \end{tikzpicture}

%% file: basics.tex
\section{Non-perturbative shift in the $3$-jet limit}
\label{sec:prelims}
	 At leading order, the final state in $e^+ e^-$ annihilation is composed of three partons, a quark, an antiquark and a gluon, forming three colour dipoles, $q\bar q$, $qg$ and $g\bar q$. To compute the non-perturbative (NP) shift to an event shape, the key ingredient is the difference between the observable evaluated with and without the gluer. Reference~\cite{Nason:2023asn} worked out the result for a multitude of observables $V$, functions of all final-state momenta. This central quantity is denoted by $h_V(\etaus,\phius,\{\tilde{p}\},k)$, where $\etaus$ ($\phius$) is the rapidity (azimuth) of the gluer $k_{\rm us}$ with respect to the emitting dipole, $\{\tilde{p}\}$ denotes the four-momenta of the $q\bar{q}$ pair, and $k$ is the momentum of the final-state perturbative gluon.

    In this work, we focus on the thrust variable, which reads
    \begin{align}
        T \equiv \frac{1}{Q}\max_{\vec{n}_T} \sum_i |\vec{p}_i \cdot \hat{n}_T|, \quad \tau = 1-T \, ,
    \end{align}
    and the NP function reads \cite{Nason:2023asn}:
	\begin{align}\label{eq:htaubasic}
		h_T(\etaus,\phius,\{\tilde{p}\},k)= \lim_{\kapus \to 0} \frac{1}{\kapus} \left(- |\vec{k}_{\rm us} \cdot \vec{n}_T | + |\vec{k}^+_{\rm us} \cdot \vec{n}_T | + |\vec{k}^-_{\rm us} \cdot \vec{n}_T |\right) \, ,
	\end{align}
	where $k^+_{\rm us}$ and $k^-_{\rm us}$ refer to the Sudakov components of $k_{\rm us}$ along the directions of the radiating dipole momenta, while $\kappa_{\rm us}$ is the gluer transverse momentum in the dipole frame. 
    %The result in eq..~\eqref{eq:htaubasic} applies generally to any $n$-parton {\em perturbative} configuration. 
	%	
	To reconstruct the gluer momentum in an arbitrary reference frame, we consider a three-parton configuration specified by $(p_1,p_2,p_3)$. Such a configuration can be explicitly constructed by applying the inverse of the on-shell kinematic map to the $(k_{\rm us},\{\tilde{p}\},k)$ configuration.
    We pick a dipole $(i,j)$, not necessarily colour-connected, and write down the Sudakov decomposition of any massless four-momentum as follows \cite{Arpino:2019ozn}:
	\begin{align}\label{eq:Sud_dipole}
		k = z^{(i)} \, p_i + z^{(j)} \, p_ j + \kappa \cos\phi \, n_{\rm in} + \kappa \sin\phi \, n_{\rm out} \, ,
	\end{align}
	where $n_{\rm in}$ and $n_{\rm out}$ are space-like vectors orthogonal to $(p_i,p_j)$. In particular,
	\begin{align}
		n_{\rm in} =\left(\cot\frac{\theta_{ij}}{2},  \frac{\vec{n}_i + \vec{n}_j}{\sin \theta_{ij}}\right) \, , \quad n_{\rm out} = \left( 0, \frac{\vec{n}_i \times \vec{n}_j}{\sin\theta_{ij}}\right) \, ,
	\end{align}
	and $\vec{n}_i = \vec{p}_i/E_i$.
	In eq.~\eqref{eq:htaubasic}, $\kappa$ is the invariant transverse momentum with respect to the radiating dipole and is given by,
	\begin{align}
		\kappa^2 = \frac{(2p_i \cdot k)(2p_j \cdot k)}{(2p_i \cdot p_j)} \, .
	\end{align}
	We stress the pair $(p_i,p_j)$ do not correspond to the physical final-state momenta, nevertheless, the gluer is softer than any perturbative emission, i.e. $k_{\rm us} \to 0$. Therefore, up to $\kapus^2$ corrections we take $(p_1,p_2,p_3) \simeq (\{\tilde{p}\},k)$.
	The Sudakov coefficients can further be expressed in terms of $(\kappa,\eta)$ as such:
	\begin{align}
		z^{(i)} = \frac{\kappa}{Q_{ij}} e^{\eta}\, , \qquad z^{(j)} = \frac{\kappa}{Q_{ij}} e^{-\eta} \, ,
	\end{align}
	where $Q^2_{ij} = 2 p_i \cdot p_j$ is the dipole invariant mass sqaured. Using the Sudakov decomposition, eq.~\eqref{eq:htaubasic} is expressed in the general form
    \begin{align}\label{eq:htaugen}
		h_T(\etaus,\phius,\{\tilde{p}\},k) = - | A e^{\etaus} + B e^{-\etaus} + C \cos\phius | + |A| e^{\etaus} + |B| e^{-\etaus} \, ,
	\end{align}
	where the various functions $(A,B,C)$ depend only on the kinematics of the $q\bar{q}g$ perturbative state. In particular, we have
	\begin{align}\label{eq:ABCexact}
		A = \frac{E_i}{Q_{ij}} \cos\theta_{iT}\, , \quad B = \frac{E_j}{Q_{ij}} \cos\theta_{jT}\, , \quad C = \frac{\cos\theta_{iT} + \cos\theta_{jT}}{\sin\theta_{ij}} \, ,
	\end{align}
	where $T$ denotes the direction of the thrust axis. 

 In a three-parton configuration, the thrust axis is aligned with the direction of the hardest parton, up to a sign ambiguity. For definiteness, we adopt the convention that it points parallel to the hardest parton. 
    The central definition of $h_T$ in eq.~\eqref{eq:htaubasic} depends not only on the choice of radiating dipole but also on the orientation of the thrust axis. This dependence is encoded in the functions $(A,B,C)$, for which we adopt the notation
    \begin{align}
    h_T^{ij/\vec{n}_T}, \quad A^{(ij)}_{\hat{n}_T}, \quad B^{(ij)}_{\hat{n}_T}, \quad C^{(ij)}_{\hat{n}_T},
    \end{align}
    where, for example, $A^{(q\bar{q})}_{q}$ denotes the case in which the thrust axis is aligned with the quark direction and the radiating dipole is $q\bar{q}$. The phase space of the $q\bar{q}g$ final state can be conveniently parametrised in terms of the Dalitz variables $(x_1,x_2)$ defined in Appendix~\ref{app:funcs}. Explicit expressions for the relevant functions in terms of $(x_1,x_2)$, corresponding to the different dipole–thrust axis combinations, are also provided in Appendix~\ref{app:funcs}.

    It is non-trivial to analytically integrate eq.~\eqref{eq:htaugen} over rapidity and azimuth, {\em viz.},
    \begin{align}\label{eq:h_average}
    \int d\etaus \, \frac{d\phius}{2\pi} \, h_T(\etaus,\phius,\{\tilde{p}\},k) &\equiv \langle h_T(\etaus,\phius,\{\tilde{p}\},k) \rangle \, ,
    \end{align}
    because the result depends on the signatures of the various functions. There are two cases to be considered:

	\paragraph{Case I: $\mathbf{sgn(B) = -\,sgn(A) = - sgn(C)}$.}
    In this case, performing the azimuthal-rapidity average in eq.~\eqref{eq:h_average} yields the following,
    \begin{align}\label{eq:h_average_secI}
		 \langle h_T(\etaus,\phius,\{\tilde{p}\},k) \rangle 
		= 4 \sqrt{|A B|} + 2 |C| \int_0^1 du \, \frac{I(u)}{\sqrt{u^2 + 4 |A B|/C^2 } } \, ,
	\end{align}
    with
	\begin{align}
		I(u) \equiv - \frac{2\, (u \arcsin(u) + \sqrt{1-u^2})}{\pi} + u \, .
	\end{align}

    In the $q\bar{q}g$ state, the current case is relevant when the thrust axis is parallel or anti-parallel to either of the momenta of the dipole ends, i.e., $p_i$ or $p_j$ in eq.~\eqref{eq:Sud_dipole}. In the former choice, we get 
    \begin{align}
        A>0, \quad B<0, \quad C>0 \, ,
    \end{align}
    while in the latter choice we have
    \begin{align}
        A<0, \quad B>0, \quad C<0 \, .
    \end{align}
    It is obvious that the final result is independent of our choice, and thus we stick with the convention of aligning the thrust axis parallel to the hardest parton. Notice that the result depends on the product $A B$, and not on the individual functions.
	\paragraph{Case II: $\mathbf{sgn(B) = \,sgn(A) = sgn(C)}$.}
	In this case we get,
    \begin{align}\label{eq:h_average_secII}
		\langle h_T(\etaus,\phius,\{\tilde{p}\},k) \rangle = -\frac{4}{\pi} |C| \left[\left(1+\frac{4 A B}{C^2}\right) K\left(1-\frac{4 A B}{C^2} \right) - 2 E\left(1-\frac{4 A B}{C^2} \right)\right]\, ,
	\end{align}
    where $K$ and $E$ are complete elliptic functions of the first and second kind respectively. Once again we observe that the result depends on the product $A B$, and not on the individual functions. This case is relevant when the thrust axis is parallel or anti-parallel to the left-over parton in the $q\bar{q}g$ state, i.e., $\hat{n}_T ||\pm \vec{p}_m$ for $m \neq i,j$. In our convention of taking the thrust axis parallel to the hardest parton, we have
	\begin{align}
		A<0 \, , \quad B < 0 \, , \quad C <0 \, .
	\end{align}
    Another useful representation of eq.~\eqref{eq:h_average_secII} is the following 1-fold integral,     
    \begin{multline}\label{eq:h_average_secII_1D}
		\langle h_T(\etaus,\phius,\{\tilde{p}\},k) \rangle = \frac2\pi |C| \,\Theta \left(1 - 4A B/C^2\right)\, \left(1- \frac{4AB}{C^2}\right) \\
        \int_0^1 du \, \left(\sqrt{\frac{u}{1-u}} - \sqrt{\frac{1-u}{u}} \right) \frac{1}{\sqrt{u + (1-u) 4AB/C^2}} \, .
	\end{multline}

%% file: recovery.tex
\section{Recovery of known results}
\label{sec:recovery}
\begin{figure}[htbp]
    \centering
    \begin{subfigure}{0.49\textwidth}
    \centering
    \includegraphics[page=1,width=1.1\linewidth]{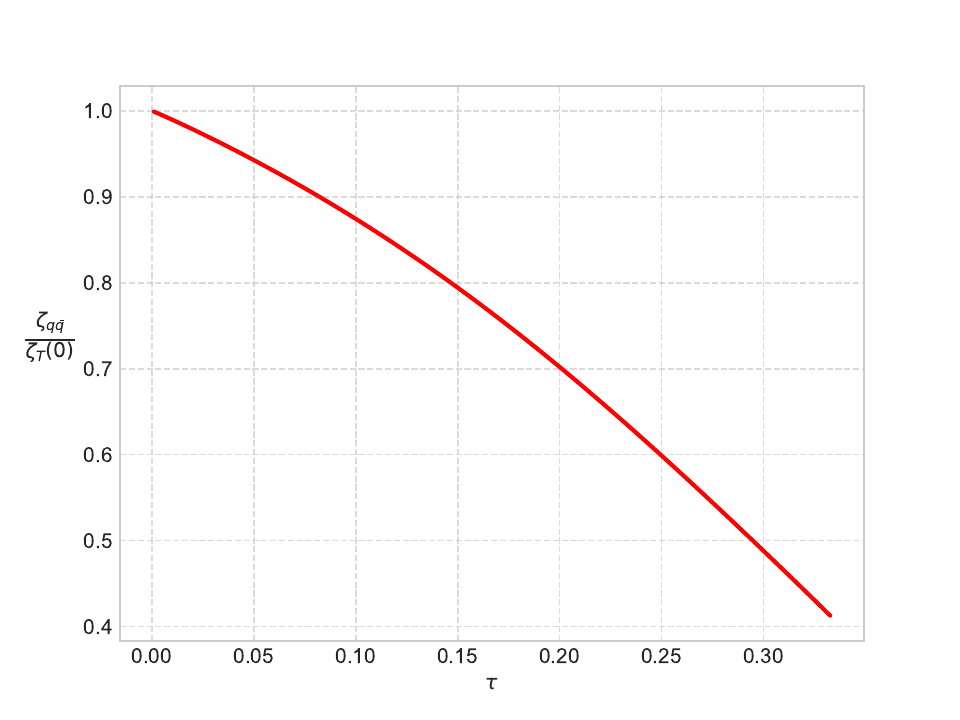}
    \subcaption{}
    \end{subfigure}
    \hfill
    \begin{subfigure}{0.49\textwidth}
    \centering
    \includegraphics[page=2,width=1.1\linewidth]{figures/zeta_qqb.pdf}
    \subcaption{}
    \end{subfigure}
    \caption{The numerical evaluation of $\zeta_{q\bar{q}}(\tau)/\zeta_T(0)$ both in linear (a) and logarithmic (b) bins.}
    \label{fig:zetaqqbar}
    \end{figure}

	In this section we employ the results in eqs.~\eqref{eq:h_average_secI} and \eqref{eq:h_average_secII} to recover the results in~\cite{Caola:2022vea}, using the thrust as a case study.
    All analytic models of hadronisation assume that power corrections are much smaller than the perturbative contribution, and can therefore be represented as a \emph{shift} in the perturbative distribution.
    %
    % Given the symmetry of the underlying $q\bar{q}g$ state, we have two cases to consider; the $q\bar{q}$ dipole and, without loss of generality, the $qg$ dipole with an appropriate factor of $2$ to account for the $\bar{q}g$ dipole.
    %
    Working under this assumption, the non-perturbative contribution to the observable cumulant, $\delta\Sigma_{\rm NP}(v)$, is given by \cite{Banfi:2023mes}:\
    \begin{equation}
    \label{eq:SigmaNP}
	   \delta\Sigma_{\rm NP}(v) = - \langle\delta V_{\rm NP}\rangle \int dZ[\{k_i\}] \, \delta\big(v-V(\{\tilde p\},\{k_i\})\big)\, ,
    \end{equation}
    where $dZ[\{k_i\}]$ is an integration measure, associated with multiple perturbative emissions and the corresponding virtual corrections, with the normalization
    \begin{equation} \label{eq:dZ-normalization}
	   \int dZ[\{k_i\}] = 1 \ \ .
    \end{equation}
    The non-perturbative shift is encoded in $\langle\delta V_{\rm NP}\rangle$,
    \begin{equation}\label{eq:avg_V_2jet}
	\langle\delta V_{\rm NP}\rangle \equiv 
	\frac{\int [dk_{\rm us}] \mathcal{M}^2_{\rm NP}(k_{\rm us}) \int dZ[\{k_i\}] \, \delta V_{\rm NP}(\{\tilde p\},k_{\rm us},\{k_i\}) \, \delta\big(v-V(\{\tilde p\},\{k_i\})\big) } {\int dZ[\{k_i\}] \, \delta\big(v-V(\{\tilde p\},\{k_i\})\big)} \, ,
    \end{equation}
    where the emission density of the gluer, $\mathcal{M}^2_{\rm NP}(k_{\rm us})$, is assumed to be uniform in rapidity and azimuth, but has an otherwise undetermined transverse momentum distribution. 
    The observable difference, $\delta V_{\rm NP}(\{\tilde p\}, k_{\rm us}, \{k_i\})$, generally depends on all perturbative emissions,
    \begin{align}
    \delta V_{\rm NP}(\{\tilde p\}, k_{\rm us}, \{k_i\}) 
    &= V(\{\tilde p\}, k_{\rm us}, \{k_i\}) 
     - V(\{\tilde p\}, \{k_i\}) \, ,
    \end{align}
    and, since we are interested in linear power corrections, we consider the limit
    \begin{equation}\label{eq:lin_V}
    \lim_{\kappa_{\rm us} \to 0} 
    \delta V_{\rm NP}(\{\tilde p\}, k_{\rm us}, \{k_i\}) 
    = \frac{\kappa_{\rm us}}{Q} \,
      h_V(\{\tilde p\}, \eta_{\rm us}, \phi_{\rm us}, \{k_i\}) \, .
    \end{equation}
    We now consider the exact $3$-jet limit, and thus expand $dZ[\{k_i\}]$ at $\mathcal{O}(\alpha_s)$:
    \begin{equation}\label{eq:dZas}
        dZ[\{k_i\}] = \frac{C_F \alpha_s}{2\pi} \left( \frac{x_1^2 + x_2^2}{(1-x_1)(1-x_2)} \right) \Theta(x_1+x_2-1)\, dx_1 dx_2 + \frac{C_F \alpha_s}{2\pi} \mathcal{V}\\
        + \mathcal{O}(\alpha_s^2) \, .
    \end{equation}
    Since virtual corrections, denoted by $\mathcal{V}$, provide just the normalisation in eq.~\eqref{eq:dZ-normalization}, they do not change the final-state kinematics, and hence contribute only at \(v = 0\), and hence do not affect the distribution for \(v > 0\) at \(\mathcal{O}(\alpha_s)\). For the gluer, we have the following emission probability:
    \begin{equation}\label{eq:gluerprob}
    [dk_{\rm us}]\mathcal{M}^2_{\rm NP}(\kus)  = \sum_{\rm dip.} C_{\rm dip.} \frac{d\kapus}{\kapus} M_{\rm NP}^2(\kapus) \,d\etaus \frac{d\phius}{2\pi} \, ,
    \end{equation}
    where the phase space is Sudakov-decomposed using the light-like momenta of the final-state colour dipoles, see eq.~\eqref{eq:Sud_dipole}, and the colour factors read:\footnote{Notice that a factor of $C_F$ is absorbed in the definition of $M_{\rm NP}^2$.}
    \begin{equation}
        C_{q\bar{q}} = 1 - \frac{C_A}{2C_F} \, , \quad C_{qg}=C_{g\bar{q}} = \frac{C_A}{2C_F} \, .
    \end{equation}
    In the notation of~\cite{Nason:2023asn}, we have:
    \begin{align}\label{eq:intro_zeta}
        \langle\delta V_{\rm NP}\rangle = \frac1Q\, \zeta_V(v) \int d\kapus \, M_{\rm NP}^2(\kapus) \, .
    \end{align}
	\paragraph{The $q\Bar{q}$ dipole.}In the region of phase space where the thrust axis aligns with the $q$ or $\bar{q}$ directions, we use the result in eq.~\eqref{eq:h_average_secI}, while eq.~\eqref{eq:h_average_secII} applies in the region where the thrust axis aligns with the gluon. We denote by $\zeta_{ij}(\tau)$ the leading-order dipole contribution to $\zeta_T(\tau)$, with the colour factor stripped off.
    Thus we compute, 
    \begin{align}\label{eq:zetaqqbar}
        \nonumber
        \zeta_{q\bar{q}}(\tau) = \left(\frac{1}{\sigma} \frac{d\sigma}{d\tau}\right)^{-1} \frac{C_F \alpha_s}{2\pi} &\int dx_1 dx_2 \left( \frac{x_1^2 + x_2^2}{(1-x_1)(1-x_2)} \right) \Theta(x_1+x_2-1) \\
        &\sum_{k=q,\bar{q},g} \langle h^{q\bar{q}/k}_T \rangle \delta(\tau-(1-\text{max}(\{x_k\})))  \, .
    \end{align}
    The plot of $\zeta_{q\bar{q}}(\tau)$ is shown in Fig.~\ref{fig:zetaqqbar} normalised to the 2-jet result $\zeta_T(0) = 2$, and is fully consistent with ref.~\cite{Caola:2021kzt}. It is evident from the log-plot that $\zeta_{q\bar{q}}(\tau)$ approaches its two-jet limit rapidly. We show in Sec.~\ref{sect:2jextract} that this feature arises because, as $\tau \to 0$, the numerator in eq.~\eqref{eq:zetaqqbar} becomes proportional to the leading-order thrust distribution.

\begin{figure}[htbp]
    \centering
    \begin{subfigure}{0.49\textwidth}
    \centering
    \includegraphics[page=1,width=1.1\linewidth]{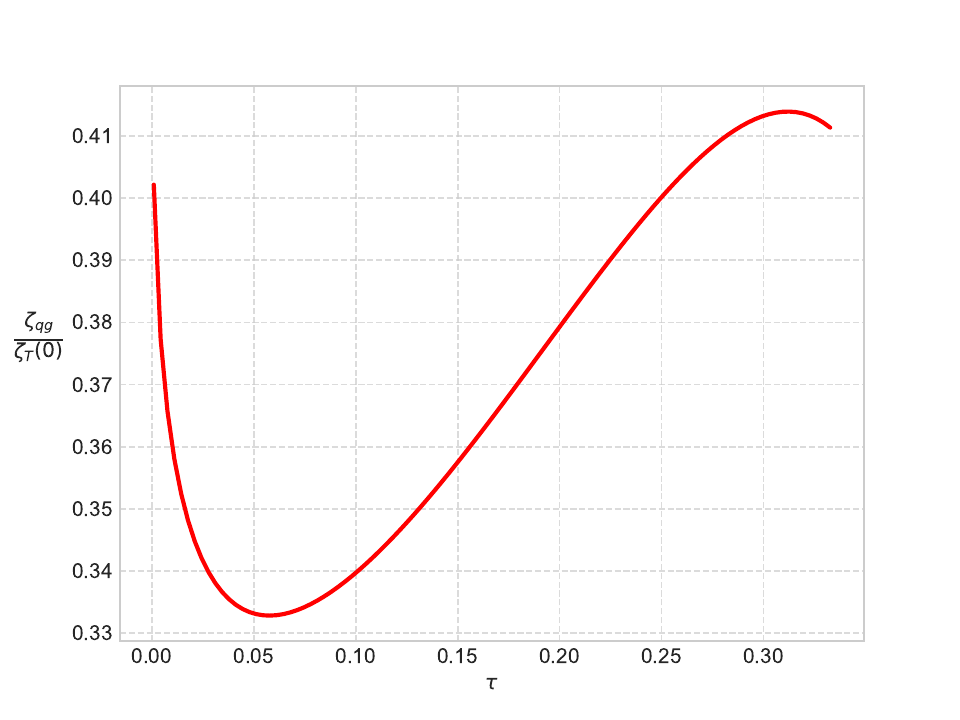}
    \subcaption{}
    \end{subfigure}
    \hfill
    \begin{subfigure}{0.49\textwidth}
    \centering
    \includegraphics[page=2,width=1.1\linewidth]{figures/zeta_qg.pdf}
    \subcaption{}
    \end{subfigure}
    \caption{The numerical evaluation of $\zeta_{qg}(\tau)/\zeta_T(0)$ both in linear (a) and logarithmic (b) bins.}
    \label{fig:zetaqg}
    \end{figure}

	%
	% \begin{figure}[h]
	% 	\centering
	% 	\includegraphics*[width=0.6\textwidth]{figures/shift_qqbar_normalized.pdf}
	% 	\caption{The contribution of the $q\Bar{q}$ dipole to the NP shift.} 
	% 	\label{fig:qqbar-normal}
	% \end{figure}
	%
	% We stress that Fig.~\ref{fig:qqbar-normal} is normalized both using the full LO thrust distribution and the 2-jet limit of the shift, {\em viz.} $\zeta(0)=2$.
	%
	\paragraph{The $qg$ dipole.}Likewise, the contribution of the $qg$ dipole can be obtained and is shown in Fig.~\ref{fig:zetaqg}. Once again, we find complete agreement with ref.~\cite{Caola:2021kzt}. It is difficult to determine from the figure the limiting behaviour near $\tau \to 0$, but we checked that indeed $\lim\limits_{\tau \to 0} \zeta_{qg}(\tau)/\zeta_T(0) = 1/2$. For this dipole, the behaviour is drastically different than that of the $q\bar{q}$ dipole. In particular, $\zeta_{qg}(\tau)$ exhibits a slow asymptotic behaviour in the limit $\tau \to 0$, which confirms that the numerator in $\zeta_{qg}(\tau)$ is {\em not} proportional to the leading-order distribution. 
	%
	% \begin{figure}[h]
	% 	\centering
	% 	\includegraphics*[width=0.6\textwidth]{figures/shift_qg_normalized.pdf}
	% 	\caption{The result of the $qg$-dipole contribution to the NP shift.} 
	% 	\label{fig:qg-normal}
	% \end{figure}
	%
	\section{Extracting the $2$-jet limit}\label{sect:2jextract}
	
	In this section, our aim is to precisely determine the 2-jet limit of the non-perturbative (NP) shift, in particular, the constant $\mathcal{O}(\alpha_s)$ terms. To this end, we study the NP shift \emph{without} normalising to the leading-order (LO) thrust distribution. Specifically, we consider the quantity:
	\begin{align}
		\bar{\zeta}_{ij}(\tau) \equiv \tau \int dx_1 dx_2 \frac{d\sigma}{dx_1 dx_2} \, \sum_{k=q,\bar{q},g} \langle  h_\tau^{ij/k} \rangle \, \delta(\tau - (1-{\rm max} \{x_k\} ) ) \, ,
	\end{align}
	where the leading-order $3$-jet differential cross section reads:
    \begin{align}\label{eq:LOsigma}
      \frac{d\sigma}{dx_1 dx_2} =  \frac{C_F \alpha_s}{2\pi} \left( \frac{x_1^2 + x_2^2}{(1-x_1)(1-x_2)} \right) \Theta(x_1+x_2-1)
    \end{align}
    To set the stage for our study, we note that $\bar{\zeta}$ has the general structure
    \begin{align}\label{eq:zetabardecomp}
	\bar{\zeta} (\tau) = \frac{C_F \alpha_s}{2\pi} \left(c_1 \ln\frac{1}{\tau} + c_0 + c(\tau)\right) \, ,
    \end{align}
    where $c(\tau) \to 0$ as $\tau \to 0$, i.e., it represents a power-suppressed correction. The logarithmic term arises from the soft-collinear region of the Born phase space, while the constant term receives contributions from both the hard-collinear and soft wide-angle regions.
	In this section, our goal is to express the non-normalized shift $\bar{\zeta}$ in the decomposition of eq.~\eqref{eq:zetabardecomp} for the different dipoles.

	The strategy we adopt is to begin by imposing the soft limit, $x_1, x_2 \to 1$, on the various functions $(A, B, C)$ as well as on the squared matrix-element. This determines the coefficient $c_1$ in eq.~\eqref{eq:zetabardecomp}, along with the soft wide-angle contribution to $c_0$. We then take the appropriate hard-collinear limit to determine the remaining piece of $c_0$. Finally, $c(\tau)$ is obtained by subtracting the logarithmic and constant terms from $\bar{\zeta}(\tau)$.
	%
	
\begin{figure}[htbp]
    \centering
    \begin{subfigure}{.49\textwidth}
    \centering
    \includegraphics[page=1,width=1.1\linewidth]{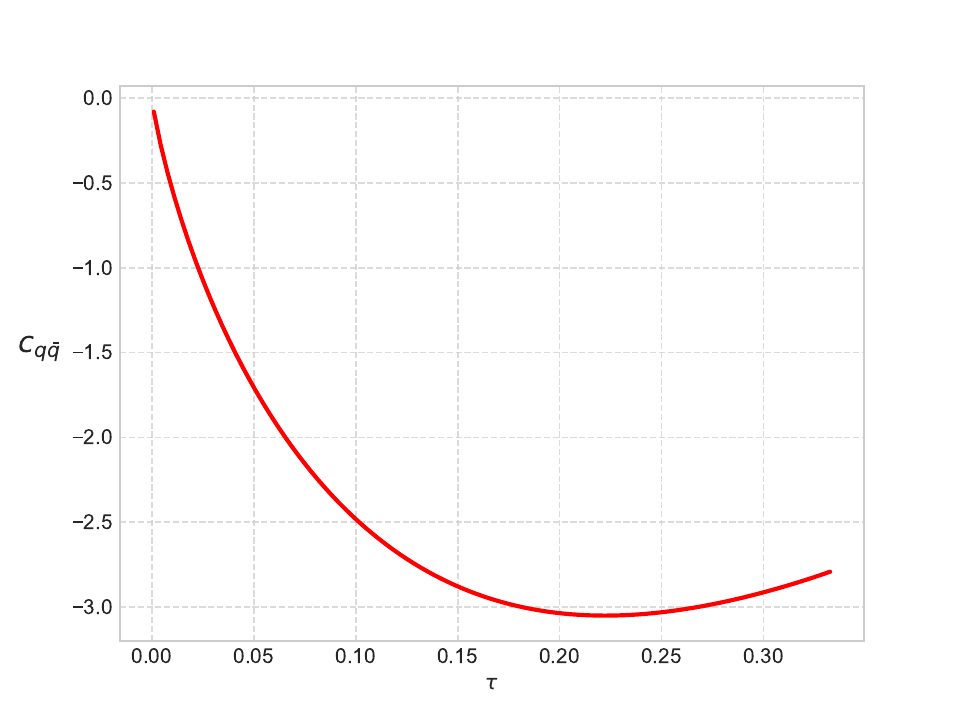}
    \subcaption{}
    \end{subfigure}
    \hfill
    \begin{subfigure}{.49\textwidth}
    \centering
    \includegraphics[page=2,width=1.1\linewidth]{figures/c_qqb.pdf}
    \subcaption{}
    \end{subfigure}
    \caption{The numerical evaluation of $c_{q\bar{q}}(\tau)$ both in linear (a) and logarithmic (b) bins.}
    \label{fig:cqqbar}
    \end{figure}

	% For future reference, we have
	% %
	% \begin{align}\label{eq:ME}
	% 	\nonumber
	% 	\frac{d\sigma^{(1)}}{dx_1 dx_2} &= \frac{C_F \alpha_s}{2\pi} \, \frac{x_1^2 + x_2^2}{(1-x_1)(1-x_2)} \Theta(x_1+x_2-1) \, , \\ 
	% 	&= \frac{C_F \alpha_s}{2\pi} \left( \frac{2}{(1-x_1)(1-x_2)} - \frac{1+x_1}{1-x_2} - \frac{1+x_2}{1-x_1} \right) \Theta(x_1+x_2-1) \, .
	% \end{align}
	%
	\paragraph{$q\Bar{q}$ dipole.}
	The 2-jet limit in this case is quite simple. First, we observe that the region of Born phase space where the thrust axis aligns with the gluon direction yields a power-suppressed contribution, i.e., it only affects $c(\tau)$.
	Without loss of generality, we consider the phase space region where the thrust axis lies along $q$. The soft limit is achieved as follows:
	\begin{align}\label{eq:avh2jqqb}
		\lim_{x_1,x_2 \to 1} A_{q}^{(q\Bar{q})} = \frac12\, , \quad \lim_{x_1,x_2 \to 1} B_{q}^{(q\Bar{q})} = -\frac12 \, , \quad \lim_{x_1,x_2 \to 1} C_{q}^{(q\Bar{q})} = 0 \, .
	\end{align}
	The constancy of the functions in the soft limit then imply that the soft region does not induce a pure constant, but only a logarithm. Thus we conclude,
	\begin{align}
		c_0^{\rm w.a.} = 0 \, .
	\end{align}
	We now move on to investigate the relevant hard-collinear limit, i.e. $x_1 \to 1$ at fixed $x_2$, and we get:
	\begin{align}
		\lim_{x_1 \to 1} |A_{q}^{(q\Bar{q})} B_{q}^{(q\Bar{q})} | = \frac14 \, , \quad \lim_{x_1 \to 1} C_{q}^{(q\Bar{q})}  = 0 \, .
	\end{align}
	We observe the interesting feature that the soft and hard-collinear limits of the functions $(A, B, C)$ coincide. This immediately allows us to write down the desired decomposition,
	\begin{align}
		\bar{\zeta}_{q\bar{q}}(\tau) = \frac{\alpha_s C_F}{2\pi} \, \left[2 \left(4 \ln\frac{1}{\tau} - 3\right) + c_{q\bar{q}}(\tau) \right]  \, ,
	\end{align}
	where the power-correction contribution is shown in Fig.~\ref{fig:cqqbar}, which vanishes by construction as $\tau \to 0$.
	\paragraph{$qg$ dipole.}
	Extracting the 2-jet limit of the $qg$-dipole contribution is less straightforward than in the $q\bar{q}$ case, but arguably more interesting, as it reveals a profound physics lesson. We are interested in both the soft and hard-collinear regions of phase space, which will be treated separately.
	%
	%\subsubsection{soft region}
	%
    
	To probe the soft limit we parameterise the phase space in terms of the perturbative gluon's transverse momentum and rapidity relative to the \( q\bar{q} \) dipole. Explicitly, we write:
	\begin{align}
		k = \frac{k_t}{\sqrt{s_{q\bar{q}}}} e^{\eta} \, p_q +\frac{k_t}{\sqrt{s_{q\bar{q}}}} e^{-\eta} \, p_{\bar{q}} + k_\perp \, , \quad k_\perp ^2 = -k_t^2\, ,
	\end{align}
	with $s_{q\bar{q}} = 2 p_q \cdot p_{\bar{q}} \sim Q^2$ in the soft limit, {\em viz.},
	\begin{align}
		1-x_1 = \frac{k_t}{Q} e^{\eta} \, ,\quad 1-x_2 = \frac{k_t}{Q} e^{-\eta} \, .
	\end{align}
	Using these variables, we recover the soft gluon emission rate from eq.~\eqref{eq:LOsigma},
	\begin{align}
		\lim_{k_t \to 0} dx_1 dx_2 \frac{2}{(1-x_1)(1-x_2)} = 2 \frac{ dk_t^2}{k_t^2} d\eta \, \Theta\left(|\eta| < \ln\frac{Q}{k_t}\right) \, .
	\end{align}
	We now express \((A, B, C)\) in terms of \(k_t\) and \(\eta\) in the soft limit \(k_t \to 0\). Aligning the thrust axis with the \(q\) direction, we obtain:
    \begin{align}\label{eq:abcqkt}
	A_{q}^{(qg)} &= \frac{1}{2} \sqrt{\frac{Q}{k_t}}\, e^{\eta/2}\,, \quad
	B_{q}^{(qg)} = \sqrt{\frac{k_t}{Q}}\, e^{\eta/2} \left( \cosh \eta - e^{-\eta} \right)\,, \quad
	C_{q}^{(qg)} = e^{\eta} \, ,
    \end{align}
    with \(\eta < 0\). We now consider the case when the thrust axis is aligned with the \(\bar{q}\) direction:
    \begin{align}\label{eq:abcqbkt}
	A_{\bar{q}}^{(qg)} &= -\frac{1}{2} \sqrt{\frac{Q}{k_t}}\, e^{\eta/2}\,, \quad
	B_{\bar{q}}^{(qg)} = \sqrt{\frac{k_t}{Q}}\, e^{\eta/2} \left( \cosh \eta - e^{\eta} \right)\,, \quad
	C_{\bar{q}}^{(qg)} = -e^{\eta} \,,
    \end{align}
    where \(\eta > 0\). An interesting feature of the expressions above is that the individual functions are not well defined in the limit \(k_t \to 0\). Nevertheless, the specific combinations that appear in \(\langle h_\tau^{ij/\vec{n}_T} \rangle\) remain finite in this limit. This observation implies that, strictly speaking, the perturbative gluon cannot be treated as soft from the outset. In other words, the expression in eq.~\eqref{eq:htaugen} does not exhibit a uniform scaling in $k_t$.

\begin{figure}[htbp]
    \centering
    \begin{subfigure}{.49\textwidth}
    \centering
    \includegraphics[width=1.1\linewidth]{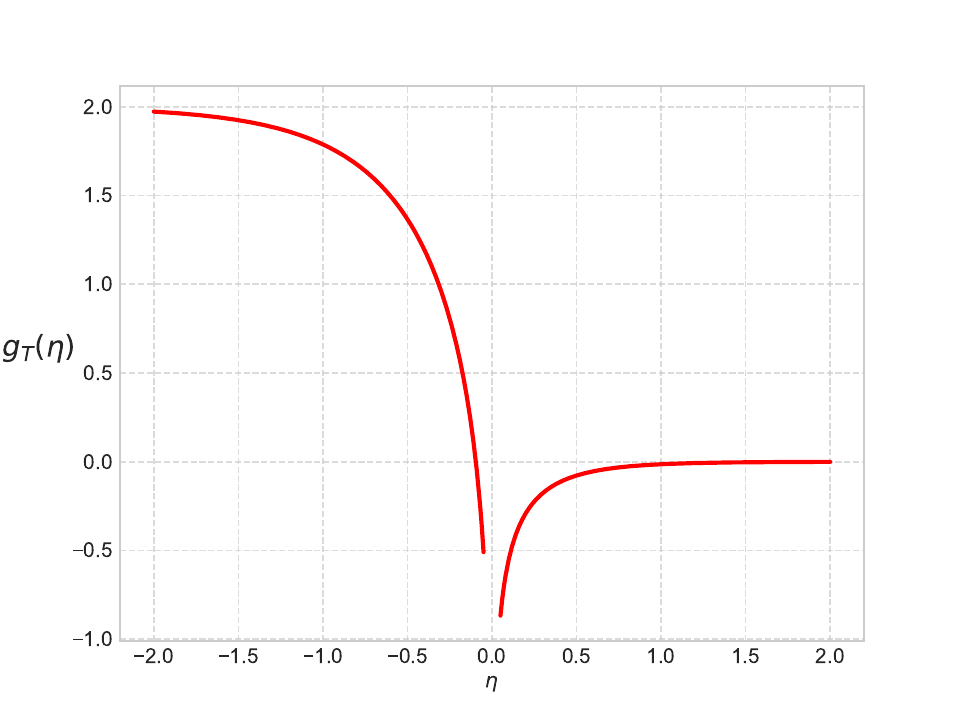}
    \subcaption{}
    \label{fig:g-eta}
    \end{subfigure}
    \hfill
    \begin{subfigure}{.49\textwidth}
    \centering
    \includegraphics[width=1.1\linewidth]{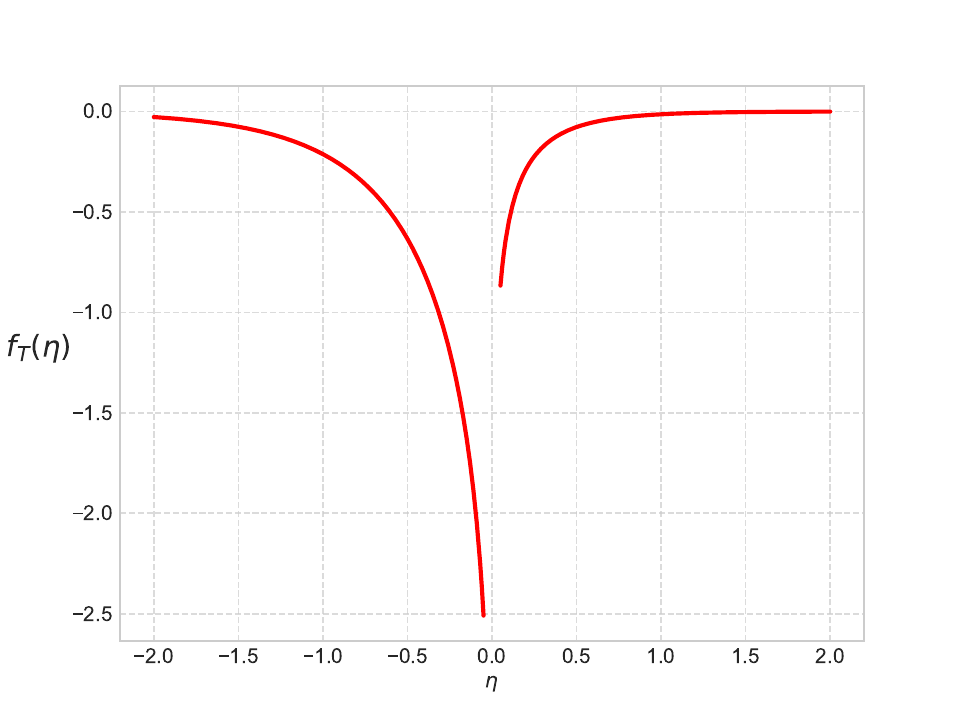}
    \subcaption{}
    \label{fig:f-eta}
    \end{subfigure}
    \caption{(a) The rapidity function $g_T(\tau)$ that determines $\bar{\zeta}_{qg}(\tau)$, (b) The rapidity function $f_T(\tau)$, which is integrable everywhere in $\eta$.}
    \end{figure}

    Therefore, we put together all these expressions to extract the $2$-jet limit of $\bar{\zeta}_{qg}$,
    \begin{align}\label{eq:zetabar-qg-basic}
        \lim_{\tau \to 0} \bar{\zeta}_{qg}(\tau) = \frac{C_F \alpha_s}{\pi} \int \frac{ dk_t^2}{k_t^2} d\eta \, \Theta\left(-\ln\frac{Q}{k_t}< \eta < \ln\frac{Q}{k_t}\right)\, \delta\left(1 - \frac{k_t}{\tau Q} \text{min}\left(e^{\eta},e^{-\eta}\right) \right) \, g_T(\tau) \, ,
        %&+ \frac{C_F \alpha_s}{\pi} \int \frac{ dk_t^2}{k_t^2} d\eta \, \Theta\left(0 < \eta <  \ln\frac{Q}{k_t}\right)\, \delta\left(1 - \frac{k_t}{\tau Q} e^{-\eta} \right) \, g_{\bar{q}}(\eta) \, ,
    \end{align}
    where the rapidity function reads:
    \begin{align}
        g_T(\tau) &= \Theta(\eta<0) \left(2 \sqrt{1 - e^{2\eta}} + 2 e^{\eta} \int_0^1 du\, \frac{I(u)}{\left(u^2 - 1 + e^{-2\eta} \right)^{1/2}} \right)\\ &+ \Theta(\eta>0)
         \,\frac{4}{\pi} e^{\eta} \bigg[\left(e^{-2\eta}-2\right)K\left(e^{-2\eta}\right) + 2 E\left(e^{-2\eta}\right)\bigg]  \, .
    \end{align}
    Due to the collinear singularity, i.e. $\vec{p}_g || \vec{p}_{\bar{q}}$, the integral of $g_T(\tau)$ can not be extended beyond the kinematic boundary. Nevertheless, we observe easily from Fig.~\ref{fig:g-eta} that the collinear limit of $g_T(\tau)$ is well defined, 
    \begin{align}
        \lim_{\eta \to -\infty } g_T(\tau) = 2 \, ,
    \end{align}
    which allows us to simply define a new function which is integrable across the whole phase space, i.e.,
    \begin{align}
        f_T(\tau) = g_T(\tau) - 2 \, \Theta(\eta<0)\, .
    \end{align}
    Fig.~\ref{fig:f-eta} shows the plot of $f_T(\tau)$ which exhibits remarkable features. Firstly, although it is divergent near $\eta =0 $, the behaviour is only logarithmic and thus integrable. Secondly, the function decays to zero exponentially fast and becomes significant only at central rapidities, i.e.\ wide-angle region of phase space. The latter feature leads us to expect that, unlike the $q\bar{q}$ case, the contribution to the constant $c_0$ in eq.~\eqref{eq:zetabardecomp} does emerge from the wide-angle region of phase space, in addition to the collinear region.

    We are now ready to rewrite eq.~\eqref{eq:zetabar-qg-basic} in a suggestive form,
    \begin{align}
    \nonumber
        \lim_{\tau \to 0} \bar{\zeta}_{qg}(\tau) &= \frac{C_F \alpha_s}{\pi} 2 \int \frac{ dk_t^2}{k_t^2} d\eta \, \Theta\left(-\ln\frac{Q}{k_t}< \eta < 0\right)\, \delta\left(1 - \frac{k_t}{\tau Q} e^{\eta} \right) \\
        &+ \frac{C_F \alpha_s}{\pi} \int \frac{ dk_t^2}{k_t^2} \int_{-\infty}^{\infty} d\eta\, f_T(\tau) \, \delta\left(1 - \frac{k_t}{\tau Q} e^{\eta} \right) \, ,
    \end{align}
    where, owing to the behaviour of $f_T(\tau)$, we extended the range of the rapidity integral beyond the kinematic boundary. In app.~\ref{app:anomdim} we show the analytic result:
    \begin{align}\label{eq:anomdim}
        \gamma \equiv \int_{-\infty}^{\infty}\!\! d\eta\, f_T(\tau) = 4 ( \ln 2 - 1), \qquad c_0^{\rm w.a.} = 4 \gamma \, .
    \end{align}

\begin{figure}[htbp]
    \centering
    \begin{subfigure}{0.49\textwidth}
    \centering
    \includegraphics[page=1,width=1.1\linewidth]{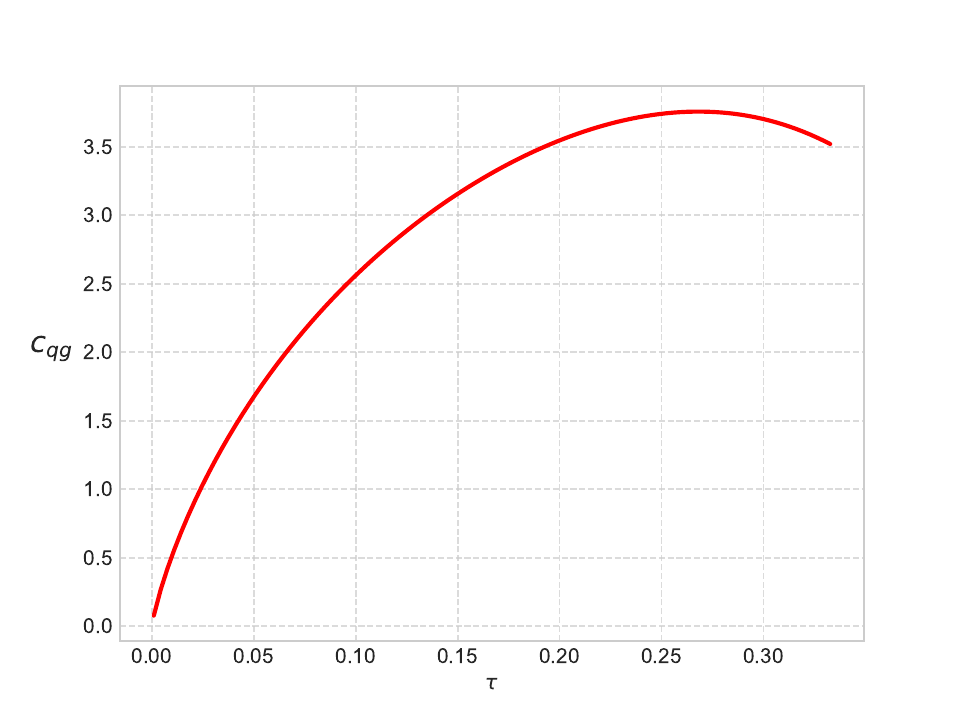}
    \subcaption{}
    \end{subfigure}
    \hfill
    \begin{subfigure}{0.49\textwidth}
    \centering
    \includegraphics[page=2,width=1.1\linewidth]{figures/c_qg.pdf}
    \subcaption{}
    \end{subfigure}
    \caption{The numerical evaluation of $c_{qg}(\tau)$ both in linear (a) and logarithmic (b) bins.}
    \label{fig:cqg}
    \end{figure}

	\paragraph{Hard-collinear region.}
	This region is simple to analyse using the expressions from the appendix. If the thrust axis aligns with the $q$ direction we have $x_1 \to 1$ at fixed $x_2$, and we easily get:
	\begin{align}
		\lim_{x_1 \to 1} |A_{q}^{(qg)} B_{q}^{(qg)} | = \frac14 \, , \quad \lim_{x_1 \to 1} C_{q}^{(q\Bar{q})}  = 0 \, ,
	\end{align}
	while when the thrust axis aligns along the $\bar{q}$ direction ($x_2 \to 1$ at fixed $x_1$) we get:
	\begin{align}
		\lim_{x_2 \to 1} \left(-\frac{4}{\pi} |C| \left[\left(1+\frac{4 A\, B}{C^2}\right) K\left(1-\frac{4 A\, B}{C^2} \right) - 2 E\left(1-\frac{4 A\, B}{C^2} \right)\right]\right)  = 0 \, .
	\end{align}
	It is straightforward to then conclude,
	\begin{align}
		c_0^{\rm h.c.} = - 3 \, .
	\end{align}
	\paragraph{Final result.}
	Now we put everything together
	\begin{align}
		\bar{\zeta}_{qg}(\tau) = \frac{\alpha_s C_F}{2\pi} \, \left(4 \ln\frac{1}{\tau} - 3 + 16 (\ln 2 -1) + c_{qg}(\tau) \right) \, ,
	\end{align}
	where the power correction function $c_{qg}(\tau)$ is shown in Fig.~\ref{fig:cqg}.

%% file: resum_dist.tex
\section{Arbitrary number of soft-collinear emissions}
\label{sec:shift-sc}
	In the previous sections, we presented a general formulation to extract the NP correction away from the two-jet limit, encapsulated by the functions $c_{q\bar{q}}$ and $c_{qg}=c_{g\bar q}$. Although we used the thrust variable as a definite example, the formulation is quite general. The functions $c_{q\bar{q}}(\tau)$ and $c_{qg}(\tau)$, represent the NP shift in the bulk of the thrust distribution, and vanish as a power of $\tau$ (possibly enhanced by logarithms of $\tau$) in the two-jet limit. In this limit, however, the value of the thrust approaches zero, leading to the usual logarithmic enhancements, i.e. $\alpha_s \ln1/\tau \sim \mathcal{O}(1)$, to all orders in perturbation theory. In this region of phase space, we must append our computation of the shift with an arbitrary number of soft-collinear emissions.

	We start by re-arranging eq.~\eqref{eq:avg_V_2jet} and introduce the computable (perturbative) average,
	\begin{equation}\label{eq:2jshift_bar}
		\bar{\zeta}_V(v, \kapus) \equiv 
		\frac{\int dZ_{\rm sc}[\{k_i\}] \, \int d\etaus \frac{d\phius}{2\pi} \, \delta V_{\rm NP}(\{\tilde p\},k_{\rm us},\{k_i\}) \, \delta\bigg(v -  V_{\rm sc}(\{\tilde p\},\{k_i\})  \bigg) } {\int dZ_{\rm sc}[\{k_i\}] \, \delta\bigg(v - V_{\rm sc}(\{\tilde p\},\{k_i\}) \bigg)} \, ,
	\end{equation}
    such that\footnote{To avoid a proliferation of notation, we use $\bar{\zeta}$ to denote the quantity defined in eq.~\eqref{eq:2jshift_bar}, although it bears no relation to the quantity introduced in Sec.~\ref{sect:2jextract}.}
    \begin{align}
        \langle \delta V_{\rm NP} \rangle = \frac1Q \int d\kapus\, M_{\rm NP}^2(\kapus) \,\bar{\zeta}_V(v, \kapus)\, .
    \end{align}
	%
    %
    %
	%, and as such the gluer does not resolve the color charge of any single perturbative gluon, i.e., the gluer is emitted from the $q\bar{q}$ pair. 
    %  Thus, in the strict 2-jet limit it suffices to use the following emission probability for the gluer
	% \begin{align}
	% 	[d\kus] \mathcal{M}^2_{\rm NP}(\kus)  =  \frac{d\kapus}{\kapus} M_{\rm NP}^2(\kapus) \,d\etaus \frac{d\phius}{2\pi} \, , 
	% \end{align}
	% where $M_{\rm NP}^2(\kapus)$ is an unknown squared matrix-element whose leading moment forms the non-perturbative parameter.
	%
	In eq.~\eqref{eq:2jshift_bar} the denominator represents the NLL resummed distribution, {\em viz.},
	\begin{align}
		\frac{d\Sigma_{\rm NLL}}{dv} = \int dZ_{\rm sc}[\{k_i\}] \, \delta\bigg(v -  V_{\rm sc}(\{\tilde p\},\{k_i\})  \bigg) \, .
	\end{align}
    At the NLL level, perturbative emissions are soft and collinear, and the phase space measure reads:
	\begin{equation}
    \label{eq:sc-measure}
		dZ_{\rm sc}[\{k_i\}] = e^{-R(v)} e^{-\int^v_{\epsilon v}\, [dk]M^2(k)}
        \sum_{n=0}^\infty \frac{1}{n!} \prod_{i=1}^{n} [dk_i] M^2(k_i) \Theta\left(V_{\rm sc}(\{\tilde p\}, k_i) -\epsilon v\right)\,,
    %    \epsilon^{\Rp(v)} \sum_{n=0}^\infty \frac{1}{n!} \prod_{i=1}^{n} \sum_{\ell_i} \Rp_{\ell_i}(v) \frac{d\zeta_i}{\zeta_i} \Theta(\zeta_i - \epsilon) \, ,
	\end{equation}
	where $[dk] M^2(k)$ is the probability distribution of soft-gluon emission,
    \begin{align}
        [dk]\, M^2(k) = \frac{C_F \alpha_s(k_t)}{\pi} \frac{dk_t^2}{k_t^2} d\eta \frac{d\phi}{2\pi} \Theta\left(|\eta| < \ln \frac{Q}{k_t}\right) \, .
    \end{align}
    In eq.~\eqref{eq:sc-measure}, $R(v)$ denotes the Sudakov radiator:
    \begin{equation}
    \label{eq:radiator}
    R(v) \equiv \int  [dk]\, M^2(k)\, \Theta\left(V_{\rm sc}(\{\tilde p\}, k_i) -v\right)\,.
    \end{equation}
    %$R' = -v \,dR/dv$ its logarithmic derivative and the subscript $\ell$ denotes a leg (or a hemisphere), see for example \cite{Banfi:2018mcq}.
    %
    Using the soft-collinear measure in eq.~\eqref{eq:sc-measure}, we obtain the known results for the NLL perturbative distributions,
    \begin{equation}
    \label{eq:dSigma-NLL}
    v\frac{d\Sigma_{\rm NLL}}{dv} = R'  e^{-R(v)}\mathcal{F}(R')\,,\qquad R'=-v\frac{dR}{dv}\, .
    \end{equation}
    %as well as the NP shifts presented in~\cite{Banfi:2023mes}. \oo{Not sure what this last sentence supposed to mean.}
    %
    
	We aim to generalize the master formula in tq.~\eqref{eq:2jshift_bar} to properly characterize the transition from the two-jet to the three-jet region. Building on the insights of the previous sections, we consider a {\em single} emission from the perturbative ensemble and allow it to probe the entirety of phase space.
    In the two-jet region, this emission is accompanied by an ensemble of soft-collinear emissions. In the hard region, we use the matching functions from the previous section to describe an $\mathcal{O}(\alpha_s)$ correction to the shift.
    If this special emission is at wide angles, the gluer resolves its colour charge, thereby leading to the dipole radiation pattern given in eq.~\eqref{eq:gluerprob}.
    Also, the contribution of any collinear emission is fully factorised. This implies that the contribution of the soft and collinear emissions accompanying the perturbative soft wide-angle gluon is also factorised. 
    
	%
	% \begin{align}
	% 	[d\kus] \mathcal{M}^2_{\rm NP}(\kus)  = \sum_{\rm dip.} C_{\rm dip.} \frac{d\kapus}{\kapus} M_{\rm NP}^2(\kapus) \,d\etaus \frac{d\phius}{2\pi} \, ,
	% \end{align}
	% where we remind that $\kapus$ refers to the invariant transverse momentum with respect to the dipole.
	%To streamline the notation, we denote by $k_{\rm us}$  the relativistic momentum of the NP gluon and by $k$ the relativistic momentum of the special perturbative emission.
    %
	% Similar to ref.~\cite{Banfi:2023mes}, we focus on observables that are linear in $\kapus$, {\em cf.} eq.
	% \begin{align}
	% 	\delta V_{\rm NP}(\{\tilde p\},\kus,k,\{k_i\}) = \frac{\kapus}{Q} h_{v}(\eta_{\rm us},\phi_{\rm us},\{\tilde p\},k,\{k_i\}) \, .
	% \end{align}
	%
    For linear observables, eq.~\eqref{eq:lin_V}, we can generalize eq.~\eqref{eq:2jshift_bar} by smoothly interpolating between the two-jet and three-jet regions of phase space. To prepare the floor for matching later on, we adjust the normalisation of the shift to the physical differential distribution as follows:
    \begin{align}\label{eq:shiftcentral}
    \frac{v}{\sigma} \frac{d\sigma}{dv}\, \bar{\zeta}_V(v, \kapus)&=  
    \int dZ_{\rm sc}[\{k_i\}] \,
    \Big(\lim_{k \to 0} \langle h_V^{q\bar{q}} \rangle\Big)
    v\,\delta\!\left(v - V_{\rm sc}(\{\tilde p\}, \{k_i\})\right)
    \nonumber \\
    &\quad 
    +
      \frac{C_A}{2C_F}
      \int [dk]\, M^2(k)
      \int dZ_{\rm sc}[\{k_i\}] \,
      \Big(\lim_{k \to 0} 
      \langle h_V^{qg} + h_V^{g\bar{q}} - h_V^{q\bar{q}} \rangle\Big)
      % \nonumber \\[-2pt]
    %&\qquad\qquad \qquad\qquad \times 
      v\,\delta\!\left(v - V_{\rm sc}(\{\tilde p\}, k, \{k_i\})\right)
    \nonumber \\[6pt]
    &\quad 
    +
      \frac{C_F\,\alpha_s(Q)}{2\pi}
      \left(
        c_{q\bar{q}}(v)
        + \frac{C_A}{2C_F}
          \big(  c_{qg}(v) + c_{g\bar{q}}(v)-c_{q\bar{q}}(v)\big)
      \right) \, .
    \end{align}
    Without any further constraints on the phase-space of the special emission,  eq.~\eqref{eq:shiftcentral} is infrared divergent because in the non-abelian contribution, the transverse momentum of the perturbative gluon is not bounded from below, due to the absence of virtual corrections sharing the same colour factor~\cite{Dokshitzer:1992ip,Dasgupta:2024znl}. However, the transverse momentum of the perturbative gluon must exceed that of the gluer, $\kappa_{\rm us}$, which effectively regulates the divergence. This makes it possible to rearrange eq.~\eqref{eq:shiftcentral} in a rather suggestive form: 
    \begin{multline}\label{eq:shift_arranged}
    \frac{v}{\sigma}\,\frac{d\sigma}{dv}\, \bar{\zeta}_V(v, \kapus)
    = 
    \int dZ_{\rm sc}[\{k_i\}] \,
    \Big(\lim_{k \to 0} \langle h_V^{q\bar{q}} \rangle\Big) v\,
    \delta\!\left(v - V_{\rm sc}(\{\tilde p\}, \{k_i\})\right) \\[6pt]
    \quad
    + 
      \frac{C_A}{2C_F}
      \int_{\kappa_{\rm us}} [dk]\, M^2(k)
      \int dZ_{\rm sc}[\{k_i\}] \,
      \Big(\lim_{k \to 0} 
      \langle h_V^{qg} + h_V^{g\bar{q}} - h_V^{q\bar{q}} \rangle\Big) \times \\[-2pt]
    \quad\quad \times 
      \Theta\!\left(v - V_{\rm sc}(\{\tilde p\}, k)\right)\,  v \,
      \delta\!\left(v - V_{\rm sc}(\{\tilde p\}, \{k_i\})\right) \\[6pt]
    \quad
    +
      \frac{C_A}{2C_F}
      \int [dk]\, M^2(k)
      \int dZ_{\rm sc}[\{k_i\}] \,
      \Big(\lim_{k \to 0}
      \langle h_V^{qg} + h_V^{g\bar{q}} - h_V^{q\bar{q}} \rangle\Big) \times \\[-2pt]
    \quad\quad \times 
      \bigg[
        v\, \delta\!\left(v - V_{\rm sc}(\{\tilde p\}, k, \{k_i\})\right)
        - \Theta\!\left(v - V_{\rm sc}(\{\tilde p\}, k)\right)\, v \,
          \delta\!\left(v - V_{\rm sc}(\{\tilde p\}, \{k_i\})\right)
      \bigg] \\[6pt]
    \quad
    +
      \frac{C_F\,\alpha_s(Q)}{2\pi}
      \left[
        c_{q\bar{q}}(v)
        + \frac{C_A}{2C_F}
          \big( c_{qg}(v) + c_{g\bar{q}}(v)-c_{q\bar{q}}(v)\big)
      \right] \, .
    \end{multline}
    eq.~\eqref{eq:shift_arranged} is one of the main results of this work. Given its complexity, we provide a summary of the main ingredients:
    \begin{itemize}
        \item The first line is nothing but the two-jet result which can be carried over from previous works for any observable \cite{Banfi:2023mes}.
        \item The second and third lines encode the effect of the soft-collinear ensemble on the configuration in which the gluer resolves the colour charge of a single soft wide-angle gluon.
        \item The last term is of $\mathcal{O}(\alpha_s)$ and can be interpreted as a hard coefficient function appended to the resummed result.
        \item The emission of a hard-collinear perturbative emission also leads to an $\mathcal{O}(\alpha_s)$ constant, and factorizes from the soft-collinear ensemble. As observed in previous sections, this is straightforward to include in our final results.
    \end{itemize}

	\subsection{The thrust and similar observables}
\begin{figure}[htbp]
    \centering
    \includegraphics[width=0.7\linewidth]{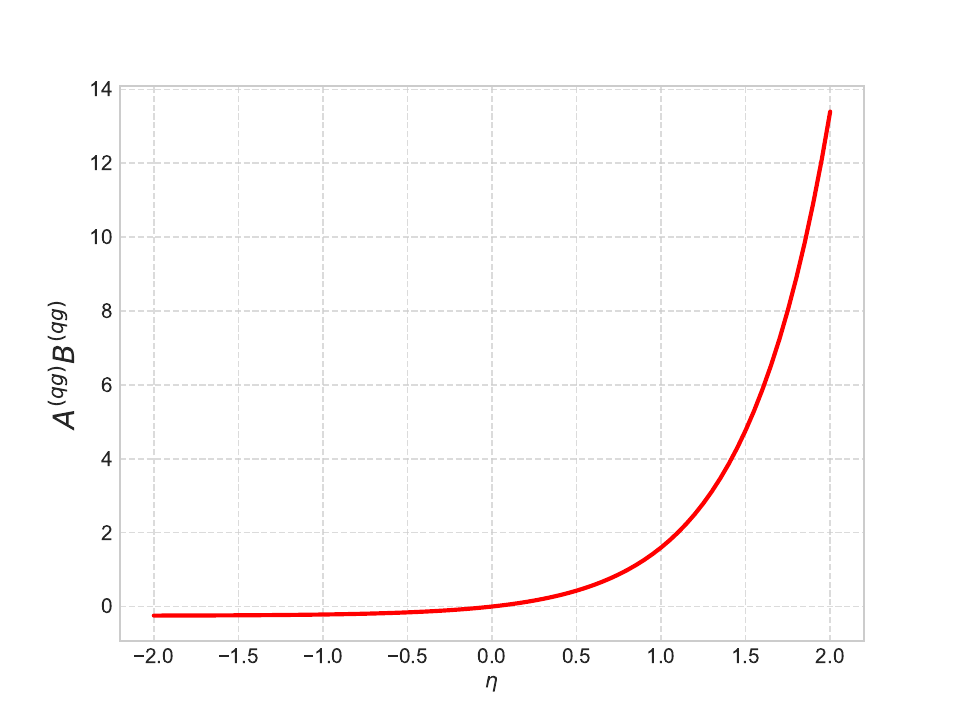}
    \caption{The plot of the product $A^{(qg)} B^{(qg)}$ as a function of rapidity.}
    \label{fig:B_function}
    \end{figure}

     Let us consider the thrust variable, and decompose the perturbative momentum $k$ in a Sudakov basis aligned with the thrust axis:
    \begin{align}\label{eq:Sudthrust}
		k = \frac{k_t}{Q} e^{\eta} \, P +\frac{k_t}{Q} e^{-\eta} \, \bar{P} + k_\perp \, , \quad k_\perp ^2 = -k_t^2\, ,
	\end{align}
	with $P = \frac{Q}{2} (1,\hat{n}_T)$ and $\bar{P} = \frac{Q}{2} (1,-\hat{n}_T)$.
    In the important region of phase space where the special perturbative emission $k$ is soft, the $q\bar{q}$ pair is nearly back-to-back, i.e., their transverse momentum is of $\mathcal{O}(k_t^2)$. Without loss of generality, we take $q(\bar{q})$ to be collinear to $P(\bar{P})$. In order to express the $(A,B,C)$ functions in terms of the transverse momentum and rapidity of the perturbative gluon, we recall eq.~\eqref{eq:ABCexact} and use eq.~\eqref{eq:Sudthrust} to find: 
    \begin{align}
	A^{(qg)} &= \frac{1}{2} \sqrt{\frac{Q}{k_t}}\, e^{\eta/2}\,, \quad
	B^{(qg)} = \sqrt{\frac{k_t}{Q}}\, e^{\eta/2} \left( \cosh \eta - e^{-\eta} \right)\,, \quad
	C^{(qg)} = e^{\eta} \, ,
    \end{align}
    and
    \begin{align}
	A^{(q\bar{q})} &= \frac{1}{2} \,, \quad
	B^{(q\bar{q})} = -\frac12\,, \quad
	C^{(q\bar{q})} = 0 \, ,
    \end{align}
	Notice that we do not label the functions by the direction of the thrust axis, since the latter is uniquely defined for the entire event, including the soft-collinear ensemble.
    To determine $\langle h_T \rangle$ in each dipole, it suffices to study the relative signs of the functions $(A,B,C)$. Both $A^{(qg)}$ and $C^{(qg)}$ are positive definite, while $B^{(qg)}$ changes sign depending on the hemisphere in which the perturbative gluon lies, i.e.\ on the sign of rapidity defined in eq.~\eqref{eq:Sudthrust}.
    For reference, the plot of $ A^{(qg)} \times B^{(qg)}$ is shown in Fig.~\ref{fig:B_function}, where it is obvious that the product changes sign with changing the sign of rapidity.
    Using the results of Sec.~\ref{sec:prelims}, it is straightforward to see that,
    \begin{align}
    \lim_{k\to 0} \,\langle 
    h_{T}^{qg} + h_{T}^{g\bar{q}} 
    - h_{T}^{q\bar{q}}  \rangle = 2 f_{T}(\eta) \, ,
    \end{align}
    where $f_{T}(\eta)$ is the function given in Fig.~\ref{fig:f-eta}.

    Therefore, for the thrust, eq.~\eqref{eq:shiftcentral} becomes: 
    \begin{multline}\label{eq:shift_thrust}
    \frac{\tau}{\sigma}\,\frac{d\sigma}{d\tau}\,
    \bar{\zeta}_T(\tau,\kapus)
    = \int dZ_{\rm sc}[\{k_i\}]\, \tau\,
    \delta \big(\tau - V_{\rm sc}(\{\tilde p\},\{k_i\})\big) \times \\
    \times
    \Bigg(
    \zeta_T(0)
  + \frac{C_A}{2C_F}
    \int_{\kappa_{\rm us}}[dk]\,
    M^2(k)\,2 f_T(\eta)\,
    \Theta \big(\tau - V_{\rm sc}(\{\tilde p\},k)\big)
    \Bigg)
    \\[3pt]
    + \frac{C_A}{2C_F}
  \int [dk]\,M^2(k)\,2f_T(\eta)
  \int dZ_{\rm sc}[\{k_i\}]\times \\
  \times
  \bigg( \tau\, 
    \delta \big(\tau - V_{\rm sc}(\{\tilde p\},k,\{k_i\})\big)
    - \Theta \big(\tau - V_{\rm sc}(\{\tilde p\},k)\big)\, \tau \,
      \delta \big(\tau - V_{\rm sc}(\{\tilde p\},\{k_i\})\big)
  \bigg)
\\[3pt]
+ \frac{C_F\,\alpha_s(Q)}{2\pi}
  \left[
    c_{q\bar{q}}(\tau)
    + \frac{C_A}{2C_F}
      \big(c_{qg}(\tau)
      + c_{g\bar{q}}(\tau)
      - c_{q\bar{q}}(\tau)\big)
  \right] .
\end{multline}
    Note that, 
    \begin{equation}
        \int dZ_{\rm sc}[\{k_i\}]\, \tau\,
    \delta \big(\tau - V_{\rm sc}(\{\tilde p\},\{k_i\})\big) = \tau \frac{d\Sigma_{\rm NLL}}{d\tau}\, ,
    \end{equation}
    where $\tau \,d\Sigma_{\rm NLL}/d\tau$ has the form in eq.~\eqref{eq:dSigma-NLL} with,
    \begin{equation}
        \mathcal{F}(R')= \frac{e^{-\gamma_E R'}}{\Gamma(1+R')}\, .
    \end{equation}
    To evaluate the second contribution, proportional to $f_T$, we reparameterize the phase space of the perturbative gluon in terms of the rapidity fraction $\xi$ and the observable fraction $\zeta$--the latter not to be confused with the non–perturbative shift. In the scale of the coupling, we set $k_t = \tau Q$, i.e.\ the typical transverse momentum suitable for a soft wide-angle gluon contributing to the observable. Thus, we get: 
    \begin{multline}\label{eq:shift_thrust_final}
    \frac{\tau}{\sigma}\frac{d\sigma}{d\tau}  \bar{\zeta}_V(\tau,\kapus)
    =  \tau \frac{d\Sigma_{\rm NLL}}{d\tau}\, \left(\zeta_T(0) + \frac{2C_A}{\pi} \int d\eta\, f_T(\eta) \int_{\kapus}^{Q \tau e^{|\eta|} } \frac{dk_t}{k_t}\, \alpha_s(k_t)  \right) \\
    + \tau \frac{d\Sigma_{\rm NLL}}{d\tau}\, \frac{2 C_A \alpha_s(\tau Q)}{\pi} \,\gamma \int_0^1 \frac{d\zeta}{\zeta} \left( (1-\zeta)^{-1+R'}  - 1 \right)  \\
     + \frac{C_F \alpha_s(Q)}{2\pi}
   \left[
     c_{q\bar{q}}(\tau) + \frac{C_A}{2C_F} 
     \left(c_{qg}(\tau)
      + c_{g\bar{q}}(\tau)
      - c_{q\bar{q}}(\tau)\right)
   \right] \, ,
    \end{multline}
    where, in the second line, we used the following result for the integral over the rapidity fraction,
    \begin{align}
        \lim_{\tau \to 0} \int_{-1}^1 d\xi \, \frac12 \ln\frac{1}{\tau \zeta}\, f_T\left(\xi \frac12 \ln\frac{1}{\tau \zeta} \right) = \gamma \, .
    \end{align}
    We observe that the effect of the soft-collinear ensemble factorizes from that of the soft wide-angle gluon, in close analogy to what takes place in the ARES approach to resummation when computing NNLL corrections \cite{Banfi:2014sua,Banfi:2018mcq}.
    The integral over $\zeta$ can be carried out analytically,
    \begin{align}\label{eq:Gofrp}
    \chi_T(R') &= \int_0^1 \frac{d\zeta}{\zeta} \left[ (1-\zeta)^{-1+R'} - 1 \right] 
    = \frac{1}{R'} - \psi^{(0)}(1 + R') - \gamma_E =\frac{1}{R'}+\frac{d}{dR'}\ln\mathcal{F}(R')\, ,
    \end{align}
    and encodes the contribution of the wide-angle gluon when inserted into the soft-collinear ensemble.

    We can finally package the final answer in the following way:
    \begin{multline}\label{eq:shift_thrust_final_1}
    \frac{\tau}{\sigma} \frac{d\sigma}{d\tau}  \bar{\zeta}_T(\tau,\kapus) 
    =  \tau \frac{d\Sigma_{\rm NLL}}{d\tau}\, \zeta_T(0) \left[1+2 C_A \gamma\, \int^{Q\tau}_{\kapus}\frac{dk_t}{k_t}\frac{\alpha_s(k_t)}{\pi}\right]
    \\ + \tau \frac{d\Sigma_{\rm NLL}}{d\tau}\, \frac{2 C_A \alpha_s(\tau Q)}{\pi}\, \left( \gamma\, \chi_T(R') + \int d\eta\, f_T(\eta) |\eta| \right)\\
     + \frac{C_F \alpha_s(Q)}{2\pi}
   \left[
     c_{q\bar{q}}(\tau) + \frac{C_A}{2C_F} 
     \left(c_{qg}(\tau)
      + c_{g\bar{q}}(\tau)
      - c_{q\bar{q}}(\tau)\right)
   \right] \, .
    \end{multline}
    The expression above mixes perturbative and non-perturbative physics since the integral extends into the region where the running coupling reaches the low scale $\kappa_{\mathrm{us}}$. However, we recognise that this term gives the anomalous dimension of the shift computed in~\cite{Dasgupta:2024znl,Farren-Colloty:2025amh}. In particular, we know from the numerical study in~\cite{Farren-Colloty:2025amh} using the PanScales shower that, in the large-$N_c$ limit, the anomalous dimension exponentiates, and that it affects the thrust distribution in the same way for all values of $\tau$. Guided by this numerical observation, we construct a bootstrap for the dependence of the non–perturbative shift on $\kappa_{\rm us}$ as follows:
    \begin{equation} 
    \label{eq:zetabar-final}
    \bar\zeta_T(\tau,\kapus) = %\frac{1}{Q}\left\langle \kapus 
     \zeta_T(\tau)\exp\left[2 C_A \gamma \int_{\kapus}^{Q}\frac{dk_t}{k_t} \frac{\alpha_s(k_t)}{\pi} \right]\, ,
    \end{equation}
    with the exponent in the above equation to be computed as follows:
    \begin{equation}
       \exp\left[2 C_A \gamma \int_{\kapus}^{Q}\frac{dk_t}{k_t} \frac{\alpha_s(k_t)}{\pi} \right] = \exp\left[2 C_A \gamma \int_{\kapus}^{\mu_I}\frac{dk_t}{k_t} \frac{\alpha_s(k_t)}{\pi} \right] \left(\frac{\alpha_s(\mu_I)}{\alpha_s( Q)}\right)^{ \frac{\gamma}{\pi \beta_0 \,\zeta_T(0)} C_A}\,.
    \end{equation}
    We also assume that eq.~\eqref{eq:zetabar-final} remains valid beyond leading colour. In the above, $\Lambda_{\rm QCD} \ll \mu_I < \tau Q$ is a {\em perturbative} infrared scale that separates the perturbative and non-perturbative regions of phase space \cite{Dokshitzer:1995zt}.
    Here, $\zeta_T(\tau)$ is the {\em perturbatively} calculable quantity,
    \begin{multline}\label{eq:shift_thrust_final_2}
    \frac{\tau}{\sigma} \frac{d\sigma}{d\tau}  \zeta_T(\tau) 
    = \left(\frac{\alpha_s(Q)}{\alpha_s(\tau Q)}\right)^{ \frac{\gamma}{\pi \beta_0 \,\zeta_T(0)} C_A}
    \bigg[  \tau \frac{d\Sigma_{\rm NLL}}{d\tau}\, \left(\zeta_T(0)
    + \frac{2 C_A \alpha_s(\tau Q)}{\pi}\, \left( \gamma \, \chi_T(R') + \int d\eta\, f_T(\eta) |\eta| \right)\right)\\
     + \frac{C_F \alpha_s(Q)}{2\pi}
   \left(
     c_{q\bar{q}}(\tau) + \frac{C_A}{2C_F} 
     \left(  c_{qg}(\tau) + c_{g\bar{q}}(\tau) - c_{q\bar{q}}(\tau)\right)
   \right) \bigg]\, .
    \end{multline}
    The above result is obtained with one-loop running -- $2\pi\beta_0 = 11C_A/6 - n_f/3$. If we substitute $\zeta_T(0)$ and $\gamma$, we recover the anomalous dimension which appeared in \cite{Dasgupta:2024znl,Farren-Colloty:2025amh}:
    \begin{align}\label{eq:anom_dim}
         \frac{C_A \gamma}{\pi \beta_0 \,\zeta_T(0)} = \frac{C_A}{\pi \beta_0} 2 (\ln 2 -1 )\, .
    \end{align} 
    We end this section by defining a recipe to determine the full non-perturbative shift for observables for which $h_V$ is independent of additional soft-collinear emissions:
    \begin{itemize}
        \item Work out the soft limit of $h_V$, e.g.\ using a construction similar to that of Sec.~\ref{sect:2jextract}.
        \item Integrate over $(\etaus,\phius)$ to determine $\langle h_V \rangle$, which subsequently fixes the function $f_V(\eta)$.
        \item The integral of $f_V(\eta)$ then yields $\gamma$ for this particular observable.
        \item Through the matching procedure outlined in Sec.~\ref{sect:2jextract}, determine the coefficient functions $c_{ij}$ for each dipole.
    \end{itemize}
    Since we have only considered the thrust, one might question the universality of the anomalous dimension in eq.~\eqref{eq:anom_dim}, since it depends explicitly on $\gamma$ and $\zeta_T(0)$. From the results of ref.~\cite{Dasgupta:2024znl}, where they consider the NP shift to the mean value of the thrust and the $C$-parameter, we infer that the ratio $\gamma/\zeta_T(0)$ is indeed observable independent, at least for observables insensitive to recoil effects.

%% file: matching.tex
\section{Matching}
\label{sec:matching}

% \begin{itemize}
%     \item package the final answer at the end of the last section DONE
%     \item talk about the 1/rp issue DONE
%     \item give away the full FO distribution DONE
%     \item give away the expression for rp and F(rp) DONE IN INTRO
%     \item give away the NLL matched distribution DONE
%     \item show modified logs and the -3
%     \item show the plot with error bands
% \end{itemize}
%
The final result in eq.~\eqref{eq:shift_thrust_final_2} is not directly suitable for phenomenological applications due to the divergent behaviour of $\chi_T(R')$ as $R' \to 0$, which corresponds to large values of~$\tau$. This behaviour is expected, since the contribution must reproduce the wide-angle constant in the $R' \to \infty$ limit-- which it does because this contribution multiplies the resummed differential distribution, which is linear in $R'$. Furthermore, the resummation must be smoothly switched off near the kinematic end-point, and the non-perturbative shift is required to asymptote to its expression in the three-jet limit.
To address these issues, we match the differential distribution, $d\sigma/d\tau$, to the $\mathcal{O}(\alpha_s)$ result and use the matched distribution to normalise the shift.

A detailed phenomenological analysis is beyond the scope of this work; therefore, here we employ a simple additive matching scheme, as follows:
\begin{align}
    \left(\frac{\tau}{\sigma}\frac{d\sigma}{d\tau}\right)_{\rm matched} &= \tau\frac{d\Sigma_{\rm NLL}}{d\tau} + \left(\frac{\tau}{\sigma}\frac{d\sigma}{d\tau}\right)_{\rm LO} - \tau \frac{d\Sigma^{\alpha_s}_{\rm NLL}}{d\tau} \, ,
\end{align} 
where $\Sigma^{\alpha_s}_{\rm NLL}$ is the $\mathcal{O}(\alpha_s)$ expansion of the resummed distribution. For the NLL differential spectrum, we adopt the following expression: 
\begin{align}
    \tau \frac{d\Sigma_{\rm NLL}}{d\tau} 
    = e^{-R(\tau)} 
    \left[
        R'_{\rm NLL} 
        - 3 C_F\frac{\,\alpha_s(\sqrt \tau Q)}{2\pi}
    \right]
    \mathcal{F}(R') \, ,
\end{align}
which formally includes NNLL contributions arising from the hard-collinear constant.
The resummation is subsequently switched off near the kinematic end-point by using modified logarithms, {\em viz.},
\begin{align}
    \ln\frac{1}{\tau} &\to \frac{1}{p}\ln\!\left(1 + \frac{1}{\tau^p} - \frac{1}{\tau_{\rm max}^p}\right), 
    \qquad 
    {\rm const} \to {\rm const} \left(1 - \left(\frac{\tau}{\tau_{\rm max}}\right)^p \right)^{1/p} \, ,
\end{align}
where $p > 0$ controls how fast resummation is turned off as the kinematic end-point is approached. We apply the same modifications to the expression on the right-hand side of eq.~\eqref{eq:shift_thrust_final_2}.
%, and is used to estimate uncertainty in our final predictions for the shift.
%

\begin{figure}[t]
\centering
\includegraphics[width=0.7\linewidth]{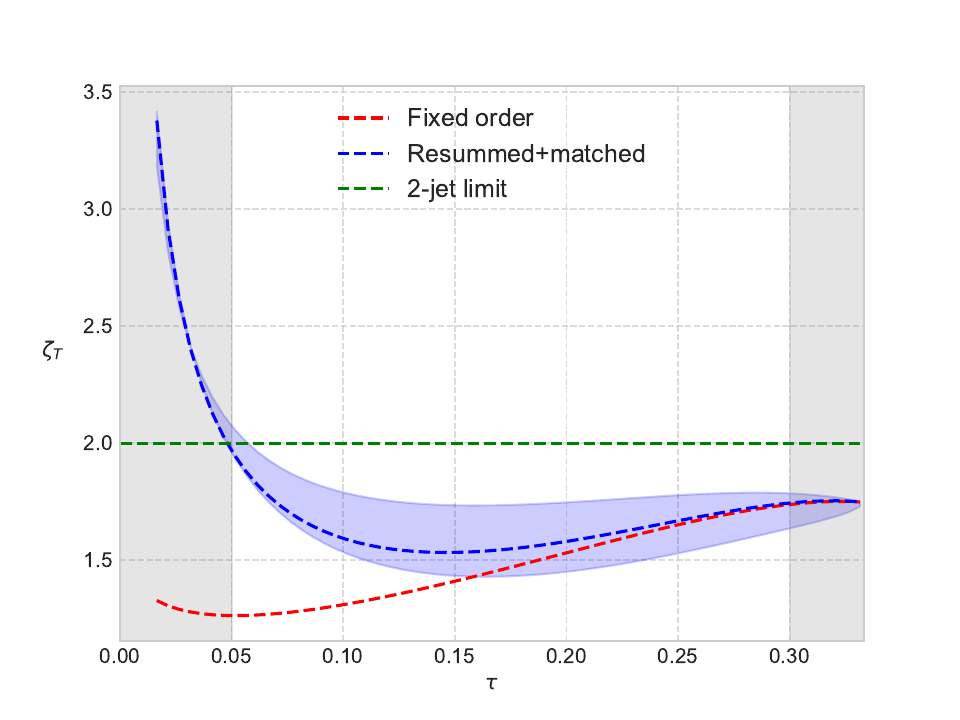}
\caption{Comparison of the quantity 
$\zeta_T(\tau)$
for the resummed+matched, fixed-order, and naive two-jet results. 
The uncertainty band on the resummed+matched curve reflects the variation of $p$ between $1/2$ and $4/3$. The grey-shaded regions are those typically excluded from the fit range \cite{Nason:2023asn}.}
\label{fig:shift}
\end{figure}

In Fig.~\ref{fig:shift}, we show the quantity $\zeta_T(\tau)$
for the matched-resummed result, the fixed-order prediction obtained in \cite{Nason:2023asn}, and the naive two-jet limit, for $Q=M_Z$ and $\alpha_s(M_Z)=0.118$.
For the matched-resummed shift, we construct a band aimed at capturing the uncertainty due to the matching procedure, by varying the $p$ between $1/2$ and $4/3$.
We observe that the matched-resummed expression for $\zeta_T(\tau)$ approaches smoothly its fixed-order counterpart for large $\tau$.
The approach to the fixed-order behaviour becomes slower for larger values of $p$. However, it is true that other theoretical uncertainties need to be considered before making any robust phenomenological predictions, such as varying the perturbative scale of the coupling.
At smaller values of $\tau$, the matched-resummed shift departs from its fixed-order expression, and is almost constant for intermediate values of $\tau$, albeit still lower than the naive two-jet limit. At further values of $\tau$, in particular close to the peak of the thrust distribution, $\tau\simeq 0.025$ at LEP1 energies, $\zeta_T(\tau)$ grows due to the behaviour of the function $\chi_T(R') \sim - \ln R'$ as $R' \to \infty$, and also the coupling $\alpha_s(\tau Q)$ approaching the Landau pole, see eq.~\eqref{eq:shift_thrust_final_2}.
Evidently, our treatment to compute the shift can not be applied in the region to the left of the peak. 
We close this section by stressing the expected impact of resummation on future determinations of $\alpha_s$, which we leave for future work.
In particular, using the fit range considered in \cite{Nason:2023asn}, i.e. $0.05 < \tau < 0.3$, the resummation clearly drives the fixed-order shift towards its na\"ive 2-jet limit showing a qualitatively different behaviour.

%% file: conclusions.tex
\section{Conclusions}
A deeper understanding of non--perturbative corrections remains essential for achieving precision in QCD, beyond its intrinsic theoretical importance in illuminating one of the most fascinating aspects of strong--interaction dynamics. Within the dispersive approach, non--perturbative effects on event--shape observables have been studied extensively, particularly in the context of $\alpha_s$ extractions. However, most previous studies have focused on restricted regions of phase space, typically the two--jet or three--jet limits, which considerably narrow the phenomenological applicability of the results and, consequently, limit the accuracy of strong--coupling determinations. In this work, we developed a general framework to characterise hadronisation corrections to event-shape variables across the entire phase space.

Although we have focused on the thrust variable, our framework is general and applies to any global event shape. Our analysis disentangles the distinct sources of $\mathcal{O}(\alpha_s)$ corrections—originating from hard--collinear and soft wide--angle emissions—and clarifies their respective colour structures. In particular, the hard--collinear contribution is proportional to $C_F$ and factorises cleanly from the soft--collinear ensemble, whereas the soft wide--angle configuration generates a genuinely new term proportional to $C_A$. The procedure for computing this latter contribution constitutes one of the central results of our work.
Furthermore, guided by recent numerical results~\cite{Farren-Colloty:2025amh}, we have proposed a bootstrap construction for the dependence of the NP shift on the ultra--soft scale $\kappa_{\rm us}$, which we conjecture to remain valid beyond leading colour. 
%This formulation offers a unified description of non--perturbative corrections across the full phase space, smoothly bridging the two--jet and multi--jet limits.
%

Our findings offer both conceptual and practical improvements over existing approaches, enabling a systematic inclusion of higher--order and multi--emission effects in the study of hadronisation corrections. Future work will extend this framework to refine the matching procedure and to cover observables beyond the thrust. Taken together, these developments pave the way towards a more universal understanding of power corrections in QCD and, ultimately, more precise extractions of the strong coupling from event--shape data.

%% file: app_ABC.tex
    \section*{Acknowledgements}

    We thank Jack Helliwell, Pier Monni, Paolo Nason, Gavin Salam and Silvia Zanoli for numerous discussions. AB acknowledges the hospitality of the Department of Physics of Royal Holloway University of London, where part of this work was performed. AB is supported by the UK Science and Technology Facility Council under the grant ST/X000796/1. BKE is supported by the Australian Research Council under grant DP220103512.
    
    \appendix
	
	\section{Expressions for various functions}\label{app:funcs}
	In this appendix we write down the expressions of the various functions $(A,B,C)$ in terms of the Born+1 phase space variables $(x_1,x_2)$, defined as $x_i\equiv 2 E_i/Q$, where $E_i$ is the energy of parton $p_i$ in the centre-of-mass frame of an $e^+e^-$ collision. Due to momentum conservation, $x_1+x_2+x_3=2$.
    For the $q\bar{q}$ dipole we have
	\begin{align}
		A^{(q\bar{q})}_q &= \frac{x_1}{2\sqrt{x_1+x_2-1}} \label{eq:Aqqbq} \, , \\
		B^{(q\bar{q})}_q &= \frac{x_2}{2\sqrt{x_1+x_2-1}} \left( 1- \frac{2(x_1+x_2-1)}{x_1 x_2} \right) \label{eq:Bqqbq}  \, , \\
		C^{(q\bar{q})}_q &= \sqrt{\frac{(1-x_1)(1-x_2)}{x_1+x_2-1}} \label{eq:Cqqbq}  \, , \\
		A^{(q\bar{q})}_g &= \frac{x_1}{2\sqrt{x_1+x_2-1}} \left(1-\frac{2(1-x_2)}{x_1(2-x_1-x_2)} \right) \, , \\
		B^{(q\bar{q})}_g &= \frac{x_2}{2\sqrt{x_1+x_2-1}} \left(1-\frac{2(1-x_1)}{x_2(2-x_1-x_2)} \right) \, , \\ \nonumber
		C^{(q\bar{q})}_g &= \frac{x_1 x_2}{\sqrt{(1-x_1)(1-x_2)(x_1+x_2-1)}} \times \\
        &\times\left(1 - \frac{1-x_2}{x_1(2-x_1-x_2)} - \frac{1-x_1}{x_2(2-x_1-x_2)} \right)\, .
	\end{align}
    The expressions for $A^{(q\bar{q})}_{\bar{q}}$, $B^{(q\bar{q})}_{\bar{q}}$ and $C^{(q\bar{q})}_{\bar{q}}$ are obtained by swapping $x_1 \leftrightarrow x_1$ in eqs.~\eqref{eq:Aqqbq}-\eqref{eq:Cqqbq}. For the $(qg)$ dipole we have the following expressions:
    \begin{align}
		A^{(qg)}_q &= \frac{x_1}{2\sqrt{1-x_2}} \, , \\
		B^{(qg)}_q &= \frac{2-x_1-x_2}{2\sqrt{1-x_2}} \left( 1 - \frac{2(1-x_2)}{x_1(2-x_1-x_2)} \right) \, , \\
		C^{(qg)}_q &= \sqrt{\frac{x_1(2-x_1-x_2)}{2(1-x_2)}} \sqrt{2- \frac{2(1-x_2)}{x_1(2-x_1-x_2)}}  \, , \\
		A^{(qg)}_{\bar{q}} &= \frac{x_1}{2\sqrt{1-x_2}} \left(1-\frac{2(x_1+x_2-1)}{x_1 x_2}\right) \, , \\
		B^{(qg)}_{\bar{q}} &= \frac{2-x_1-x_2}{2\sqrt{1-x_2}} \left(1-\frac{2(1-x_1)}{x_2(2-x_1-x_2)}\right) \, , \\ \nonumber
		C^{(qg)}_{\bar{q}} &= \sqrt{\frac{x_1(2-x_1-x_2)}{1-x_2}} \left(1-\frac{(1-x_2)}{x_1(2-x_1-x_2)}\right)^{-1/2} \times \\
        &\times\left(1 - \frac{1-x_1}{x_2(2-x_1-x_2)} - \frac{x_1+x_2-1}{x_1 x_2} \right)\, , \\
        A^{(qg)}_{g} &= \frac{x_1}{2\sqrt{1-x_2}} \left(1 - \frac{2(1-x_2)}{x_1(2-x_1-x_2)} \right) \, , \\
        B^{(qg)}_{g} &= \frac{2-x_1-x_2}{2\sqrt{1-x_2}} \, , \\ 
        C^{(qg)}_{g} &=  \sqrt{\frac{x_1(2-x_1-x_2)}{1-x_2}} \sqrt{1-\frac{(1-x_2)}{x_1(2-x_1-x_2)}} \, . 
	\end{align}
    The relevant functions for the $(g\bar{q})$ dipole are obtained by simultaneously interchanging $q \leftrightarrow \bar{q}$ and $x_1 \leftrightarrow x_2$.

    \section{The anomalous dimension}\label{app:anomdim}

    In this appendix, we detail the evaluation of eq.~\eqref{eq:anomdim}, making use of the one-fold representation given in eq.~\eqref{eq:h_average_secII_1D}. After interchanging the order of integration, the rapidity integral yields:
    \begin{align}
        \gamma =\int_0^1 du \left(2 \frac{f(u)}{1+u} + 2 \ln 2 -  \frac{4 (1-2u)}{\pi \sqrt{u} (1-u)}\arctan \left(\frac{\sqrt{u}-1}{\sqrt{1-u}} \right)  \right) \, ,
    \end{align}
    which can easily be evaluated using the transformation $u \to x^2$.

%% file: hadronisation.bib
@article{Dasgupta:2024znl,
    author = "Dasgupta, Mrinal and Hounat, Farid",
    title = "{Exploring soft anomalous dimensions for 1/Q power corrections}",
    eprint = "2411.16867",
    archivePrefix = "arXiv",
    primaryClass = "hep-ph",
    doi = "10.1007/JHEP09(2025)060",
    journal = "JHEP",
    volume = "09",
    pages = "060",
    year = "2025"
}

@article{Altarelli:1995kz,
    author = "Altarelli, Guido",
    title = "{Introduction to renormalons}",
    booktitle = "{5th Hellenic School and Workshops on Elementary Particle Physics}",
    reportNumber = "CERN-TH-95-309",
    pages = "221--236",
    year = "1996"
}

@article{Luisoni:2015xha,
    author = "Luisoni, Gionata and Marzani, Simone",
    title = "{QCD resummation for hadronic final states}",
    eprint = "1505.04084",
    archivePrefix = "arXiv",
    primaryClass = "hep-ph",
    reportNumber = "MIT-CTP-4672, MPP-2015-102",
    doi = "10.1088/0954-3899/42/10/103101",
    journal = "J. Phys. G",
    volume = "42",
    number = "10",
    pages = "103101",
    year = "2015"
}

@article{Andersson:1983ia,
    author = "Andersson, Bo and Gustafson, G. and Ingelman, G. and Sjostrand, T.",
    title = "{Parton Fragmentation and String Dynamics}",
    reportNumber = "LU-TP-83-10",
    doi = "10.1016/0370-1573(83)90080-7",
    journal = "Phys. Rept.",
    volume = "97",
    pages = "31--145",
    year = "1983"
}

@article{Webber:1983if,
    author = "Webber, B. R.",
    title = "{A QCD Model for Jet Fragmentation Including Soft Gluon Interference}",
    reportNumber = "CERN-TH-3713",
    doi = "10.1016/0550-3213(84)90333-X",
    journal = "Nucl. Phys. B",
    volume = "238",
    pages = "492--528",
    year = "1984"
}

@article{Bell:2023dqs,
    author = "Bell, Guido and Lee, Christopher and Makris, Yiannis and Talbert, Jim and Yan, Bin",
    title = "{Effects of renormalon scheme and perturbative scale choices on determinations of the strong coupling from e+e- event shapes}",
    eprint = "2311.03990",
    archivePrefix = "arXiv",
    primaryClass = "hep-ph",
    doi = "10.1103/PhysRevD.109.094008",
    journal = "Phys. Rev. D",
    volume = "109",
    number = "9",
    pages = "094008",
    year = "2024"
}

@article{Nason:2025qbx,
    author = "Nason, Paolo and Zanderighi, Giulia",
    title = "{Fits of {\ensuremath{\alpha}}$_{s}$ from event-shapes in the three-jet region: extension to all energies}",
    eprint = "2501.18173",
    archivePrefix = "arXiv",
    primaryClass = "hep-ph",
    reportNumber = "MPP-2025-9",
    doi = "10.1007/JHEP06(2025)200",
    journal = "JHEP",
    volume = "06",
    pages = "200",
    year = "2025"
}

@article{Korchemsky:1999kt,
    author = "Korchemsky, Gregory P. and Sterman, George F.",
    title = "{Power corrections to event shapes and factorization}",
    eprint = "hep-ph/9902341",
    archivePrefix = "arXiv",
    reportNumber = "ITP-SB-98-73, LPT-ORSAY-98-80",
    doi = "10.1016/S0550-3213(99)00308-9",
    journal = "Nucl. Phys. B",
    volume = "555",
    pages = "335--351",
    year = "1999"
}

@article{Farren-Colloty:2025amh,
    author = "Farren-Colloty, Casey and Helliwell, Jack and Patel, Rtvik and Salam, Gavin P. and Zanoli, Silvia",
    title = "{Anomalous scaling of linear power corrections}",
    eprint = "2507.18696",
    archivePrefix = "arXiv",
    primaryClass = "hep-ph",
    reportNumber = "OUTP-25-02P",
    month = "7",
    year = "2025"
}

@article{Nason:2023asn,
    author = "Nason, Paolo and Zanderighi, Giulia",
    title = "{Fits of $\alpha_s$ using power corrections in the three-jet region}",
    eprint = "2301.03607",
    archivePrefix = "arXiv",
    primaryClass = "hep-ph",
    month = "1",
    year = "2023"
}

@article{Banfi:2023mes,
    author = "Banfi, Andrea and El-Menoufi, Basem Kamal and Wood, Ryan",
    title = "{Interplay between perturbative and non-perturbative effects with the ARES method}",
    eprint = "2303.01534",
    archivePrefix = "arXiv",
    primaryClass = "hep-ph",
    doi = "10.1007/JHEP08(2023)221",
    journal = "JHEP",
    volume = "08",
    pages = "221",
    year = "2023"
}

@article{Arpino:2019ozn,
    author = "Arpino, Luke and Banfi, Andrea and El-Menoufi, Basem Kamal",
    title = "{Near-to-planar three-jet events at NNLL accuracy}",
    eprint = "1912.09341",
    archivePrefix = "arXiv",
    primaryClass = "hep-ph",
    doi = "10.1007/JHEP07(2020)171",
    journal = "JHEP",
    volume = "07",
    pages = "171",
    year = "2020"
}

@article{Benitez:2024nav,
    author = "Benitez, Miguel A. and Hoang, Andre H. and Mateu, Vicent and Stewart, Iain W. and Vita, Gherardo",
    title = "{On determining {\ensuremath{\alpha}}$_{s}$(m$_{Z}$) from dijets in e$^{+}$e$^{-}$ thrust}",
    eprint = "2412.15164",
    archivePrefix = "arXiv",
    primaryClass = "hep-ph",
    reportNumber = "MIT-CTP 5746, CERN-TH-2024-142, UWThPh2024-8",
    doi = "10.1007/JHEP07(2025)249",
    journal = "JHEP",
    volume = "07",
    pages = "249",
    year = "2025"
}

@article{Benitez:2025vsp,
    author = "Benitez, Miguel A. and Bhattacharya, Arindam and Hoang, Andre H. and Mateu, Vicent and Schwartz, Matthew D. and Stewart, Iain W. and Zhang, Xiaoyuan",
    title = "{A Precise Determination of $\alpha_s$ from the Heavy Jet Mass Distribution}",
    eprint = "2502.12253",
    archivePrefix = "arXiv",
    primaryClass = "hep-ph",
    reportNumber = "MIT-CTP 5840, UWTHPH 2025-7",
    month = "2",
    year = "2025"
}

@article{Gehrmann:2012sc,
    author = "Gehrmann, Thomas and Luisoni, Gionata and Monni, Pier Francesco",
    title = "{Power corrections in the dispersive model for a determination of the strong coupling constant from the thrust distribution}",
    eprint = "1210.6945",
    archivePrefix = "arXiv",
    primaryClass = "hep-ph",
    reportNumber = "ZU-TH-23-12, IPPP-12-77, DCPT-12-154, MPP-2012-143",
    doi = "10.1140/epjc/s10052-012-2265-x",
    journal = "Eur. Phys. J. C",
    volume = "73",
    number = "1",
    pages = "2265",
    year = "2013"
}

@article{Dokshitzer:1992ip,
    author = "Dokshitzer, Yuri L. and Marchesini, G. and Oriani, G.",
    title = "{Measuring color flows in hard processes: Beyond leading order}",
    reportNumber = "LU-TP-92-19, UPRF-92-330",
    doi = "10.1016/0550-3213(92)90211-S",
    journal = "Nucl. Phys. B",
    volume = "387",
    pages = "675--714",
    year = "1992"
}

@article{PhysRevD.110.030001,
  title = {Review of Particle Physics},
  author = {S.~Navas et. al.},
  collaboration = {Particle Data Group Collaboration},
  journal = {Phys. Rev. D},
  volume = {110},
  issue = {3},
  pages = {030001},
  numpages = {5},
  year = {2024},
  month = {Aug},
  publisher = {American Physical Society},
  doi = {10.1103/PhysRevD.110.030001},
  url = {https://link.aps.org/doi/10.1103/PhysRevD.110.030001}
}

@article{Dokshitzer:1995zt,
    author = "Dokshitzer, Yuri L. and Webber, B. R.",
    title = "{Calculation of power corrections to hadronic event shapes}",
    eprint = "hep-ph/9504219",
    archivePrefix = "arXiv",
    reportNumber = "CAVENDISH-HEP-95-2, LU-TP-95-8",
    doi = "10.1016/0370-2693(95)00548-Y",
    journal = "Phys. Lett. B",
    volume = "352",
    pages = "451--455",
    year = "1995"
}

@article{Akhoury:1995fb,
    author = "Akhoury, R. and Zakharov, Valentin I.",
    title = "{Leading power corrections in QCD: From renormalons to phenomenology}",
    eprint = "hep-ph/9507253",
    archivePrefix = "arXiv",
    reportNumber = "UM-TH-95-19",
    doi = "10.1016/0550-3213(96)00056-9",
    journal = "Nucl. Phys. B",
    volume = "465",
    pages = "295--314",
    year = "1996"
}

@article{Nason:1995np,
    author = "Nason, Paolo and Seymour, Michael H.",
    title = "{Infrared renormalons and power suppressed effects in
                  $e^+ e^-$ jet events}",
    eprint = "hep-ph/9506317",
    archivePrefix = "arXiv",
    reportNumber = "CERN-TH-95-150, IFUM-507-FT",
    doi = "10.1016/0550-3213(95)00461-Z",
    journal = "Nucl. Phys. B",
    volume = "454",
    pages = "291--312",
    year = "1995"
}

@article{Dokshitzer:1995qm,
    author = "Dokshitzer, Yuri L. and Marchesini, G. and Webber, B. R.",
    title = "{Dispersive approach to power behaved contributions in QCD hard processes}",
    eprint = "hep-ph/9512336",
    archivePrefix = "arXiv",
    reportNumber = "CERN-TH-95-281, CAVENDISH-HEP-95-12",
    doi = "10.1016/0550-3213(96)00155-1",
    journal = "Nucl. Phys. B",
    volume = "469",
    pages = "93--142",
    year = "1996"
}

@article{Beneke:1998ui,
    author = "Beneke, M.",
    title = "{Renormalons}",
    eprint = "hep-ph/9807443",
    archivePrefix = "arXiv",
    reportNumber = "CERN-TH-98-233",
    doi = "10.1016/S0370-1573(98)00130-6",
    journal = "Phys. Rept.",
    volume = "317",
    pages = "1--142",
    year = "1999"
}

@article{Banfi:2014sua,
    author = "Banfi, Andrea and McAslan, Heather and Monni, Pier Francesco and Zanderighi, Giulia",
    title = "{A general method for the resummation of event-shape distributions in $e^+ e^-$ annihilation}",
    eprint = "1412.2126",
    archivePrefix = "arXiv",
    primaryClass = "hep-ph",
    reportNumber = "OUTP-14-18P",
    doi = "10.1007/JHEP05(2015)102",
    journal = "JHEP",
    volume = "05",
    pages = "102",
    year = "2015"
}

@article{Banfi:2018mcq,
    author = "Banfi, Andrea and El-Menoufi, Basem Kamal and Monni, Pier Francesco",
    title = "{The Sudakov radiator for jet observables and the soft physical coupling}",
    eprint = "1807.11487",
    archivePrefix = "arXiv",
    primaryClass = "hep-ph",
    doi = "10.1007/JHEP01(2019)083",
    journal = "JHEP",
    volume = "01",
    pages = "083",
    year = "2019"
}

@article{Dasgupta:1999mb,
    author = "Dasgupta, Mrinal and Magnea, Lorenzo and Smye, Graham",
    title = "{Universality of 1/Q corrections revisited}",
    eprint = "hep-ph/9911316",
    archivePrefix = "arXiv",
    reportNumber = "BICOCCA-FT-99-34, DFTT-56-99, CAVENDISH-HEP-99-13",
    doi = "10.1088/1126-6708/1999/11/025",
    journal = "JHEP",
    volume = "11",
    pages = "025",
    year = "1999"
}

@article{Hoang:2015hka,
    author = "Hoang, Andr\'e H. and Kolodrubetz, Daniel W. and Mateu, Vicent and Stewart, Iain W.",
    title = "{Precise determination of $\alpha_s$ from the $C$-parameter distribution}",
    eprint = "1501.04111",
    archivePrefix = "arXiv",
    primaryClass = "hep-ph",
    reportNumber = "UWTHPH-2015-1, MIT-CTP-4630, LPN14-128",
    doi = "10.1103/PhysRevD.91.094018",
    journal = "Phys. Rev. D",
    volume = "91",
    number = "9",
    pages = "094018",
    year = "2015"
}

@article{Caola:2021kzt,
    author = "Caola, Fabrizio and Ferrario Ravasio, Silvia and Limatola, Giovanni and Melnikov, Kirill and Nason, Paolo",
    title = "{On linear power corrections in certain collider observables}",
    eprint = "2108.08897",
    archivePrefix = "arXiv",
    primaryClass = "hep-ph",
    reportNumber = "OUTP-21-21P, TTP21-026, P3H-21-056",
    doi = "10.1007/JHEP01(2022)093",
    journal = "JHEP",
    volume = "01",
    pages = "093",
    year = "2022"
}

@article{Caola:2022vea,
    author = "Caola, Fabrizio and Ferrario Ravasio, Silvia and Limatola, Giovanni and Melnikov, Kirill and Nason, Paolo and Ozcelik, Melih Arslan",
    title = "{Linear power corrections to $e^+e^-$ shape variables in the three-jet region}",
    eprint = "2204.02247",
    archivePrefix = "arXiv",
    primaryClass = "hep-ph",
    reportNumber = "OUTP-22-04P, TTP22-022, P3H-22-036, MPP-2022-36",
    month = "4",
    year = "2022"
}

@article{Dokshitzer:1997ew,
    author = "Dokshitzer, Yuri L. and Webber, B. R.",
    title = "{Power corrections to event shape distributions}",
    eprint = "hep-ph/9704298",
    archivePrefix = "arXiv",
    reportNumber = "CAVENDISH-HEP-97-2",
    doi = "10.1016/S0370-2693(97)00573-X",
    journal = "Phys. Lett. B",
    volume = "404",
    pages = "321--327",
    year = "1997"
}

@article{Dokshitzer:1997iz,
    author = "Dokshitzer, Yuri L. and Lucenti, A. and Marchesini, G. and Salam, G. P.",
    title = "{Universality of 1/Q corrections to jet-shape observables rescued}",
    eprint = "hep-ph/9707532",
    archivePrefix = "arXiv",
    reportNumber = "IFUM-573-FT",
    doi = "10.1016/S0550-3213(97)00650-0",
    journal = "Nucl. Phys. B",
    volume = "511",
    pages = "396--418",
    year = "1998",
    note = "[Erratum: Nucl.Phys.B 593, 729--730 (2001)]"
}
